\documentclass{aa}  

\usepackage{graphicx}
\usepackage{txfonts}
\usepackage{hyperref}
\begin{document}

   \title{A new class of dark matter-free dwarf galaxies?}

   \subtitle{I. Clues from FCC 224, NGC 1052-DF2 and NGC 1052-DF4}

   \author{Maria Luisa {Buzzo}\inst{1,2}\thanks{E-mail: marialuisa.buzzo@eso.org, lgomesbuzzo@swin.edu.au} \and
Duncan A. Forbes\inst{2} \and
Aaron J. Romanowsky\inst{3,4} \and
Lydia Haacke\inst{2} \and
Jonah S. Gannon\inst{2} \and
Yimeng Tang\inst{4} \and
Michael Hilker\inst{1} \and
Anna {Ferré-Mateu}\inst{5,6,2} \and
Steven R. Janssens\inst{7} \and
Jean P. Brodie\inst{2,8} \and
Lucas M. Valenzuela\inst{9}
          }

\institute{European Southern Observatory, Karl-Schwarzschild-Strasse 2, 85748 Garching bei M\"unchen, Germany 
\and Centre for Astrophysics and Supercomputing, Swinburne University, John Street, Hawthorn VIC 3122, Australia
\and Department of Physics and Astronomy, San José State University, One Washington Square, San Jose, CA 95192, USA 
\and Department of Astronomy \& Astrophysics, University of California Santa Cruz, 1156 High Street, Santa Cruz, CA 95064, USA 
\and Instituto Astrofisica de Canarias, Av. Via Lactea s/n, E38205 La Laguna, Spain 
\and Departamento de Astrofisica, Universidad de La Laguna, E-38200, La Laguna, Tenerife, Spain 
\and Dragonfly Focused Research Organization, 150 Washington Avenue, Santa Fe, NM 87501, USA
\and University of California Observatories, 1156 High Street, Santa Cruz, CA 95064, USA 
\and Universit\"ats-Sternwarte, Fakult\"at f\"ur Physik, Ludwig-Maximilians-Universit\"at M\"unchen, Scheinerstr. 1, 81679 M\"unchen, Germany
}

   \date{Received 19/12/2024; accepted 06/02/2025}

  \abstract
{The discovery of quiescent, dark matter (DM)-deficient ultra-diffuse galaxies (UDGs) with overluminous globular clusters (GCs) has challenged galaxy formation models within the Lambda Cold Dark Matter ($\Lambda$CDM) cosmological paradigm. Previously, such galaxies were only identified in the NGC~1052 group, raising the possibility that they are the result of unique, group-specific processes, and limiting their broader significance. The recent identification of FCC~224, a putative DM-deficient UDG on the outskirts of the Fornax Cluster, suggests that such galaxies are not confined to the NGC~1052 group but rather represent a broader phenomenon.}
{We aim to investigate the DM content of FCC~224 and to explore its similarities to the DM-free dwarfs in the NGC~1052 group, DF2 and DF4, to determine whether or not it belongs to the same class of DM-deficient UDGs.}
{We use high-resolution Keck Cosmic Web Imager (KCWI) spectroscopy to study the kinematics, stellar populations, and GC system of FCC~224, enabling direct comparisons with DF2 and DF4.}
{We find that FCC~224 is also DM-deficient and exhibits a distinct set of traits shared with DF2 and DF4, including slow and prolate rotation, quiescence in low-density environments, coeval formation of stars and GCs, flat stellar population gradients, a top-heavy GC luminosity function, and monochromatic GCs.}
{These shared characteristics signal the existence of a previously unrecognised class of DM-deficient dwarf galaxies. This diagnostic framework provides a means of identifying additional examples and raises new questions for galaxy formation models within $\Lambda$CDM cosmology.}


   \maketitle
%

\section{Introduction}

The $\Lambda$CDM cosmological model has long been the cornerstone of our understanding of galaxy formation and evolution. Central to this paradigm is the idea that dark matter (DM) is an essential component of all galaxies, providing the gravitational framework necessary for their formation, stability and evolution. The discovery of NGC~1052-DF2 and NGC~1052-DF4 (hereafter DF2 and DF4), two quiescent ultra-diffuse galaxies (UDGs) exhibiting a significant lack of DM in their inner regions, has raised questions about the validity of galaxy formation models within the accepted paradigm. These galaxies challenge the conventional understanding of how galaxies form and evolve within $\Lambda$CDM \citep{Haslbauer_19b} and the presumed relationship between galaxies and their DM halos \citep{vanDokkum_18,vanDokkum_19b}.

DF2 and DF4 possess many unusual characteristics, the most significant being their apparent total lack of DM in their inner regions. These features have sparked numerous theories about their origins \citep[e.g.,][]{Ogiya_18,Ogiya_22,Moreno_22,Jackson_21,Ivleva_24,Shin_20,Lee_21,Lee_24}. Among these is the `bullet-dwarf' scenario proposed by \citet{Silk_19} and \citet{vanDokkum_22}, wherein high velocity interactions separate dark and baryonic matter while generating the intense pressure needed to form unusually bright globular clusters (GCs). DF2 and DF4 indeed host strikingly overluminous GCs, with a top-heavy GC luminosity function (GCLF) that sets them apart from normal dwarf galaxies \citep{vanDokkum_18,vanDokkum_19b,Shen_21a,vanDokkum_22b}. The prediction from this scenario is that eight `bullet dwarf' interactions producing DM-deficient UDGs will have occurred in a 20 Mpc volume \citep{vanDokkum_22}. Galaxies formed through this process would likely constitute a new class of DM-deficient dwarfs. However, until now, such galaxies had only been identified in association with the NGC~1052 group, leaving the broader relevance of this model uncertain.

Within this context, FCC 224 -- a galaxy that was long catalogued \citep{Ferguson_89} but whose potential significance remained hitherto unrecognised-- has resurfaced as a pivotal discovery in a new catalogue of low-surface brightness galaxies \citep{Tanoglidis_21}, where its striking similarities to the DM-deficient galaxies in the NGC~1052 group, DF2 and DF4, began to draw attention. The first indication of a connection between FCC~224 and the DM-deficient galaxies in the NGC~1052 group came from its population of unusually bright GCs, a feature that prompted renewed interest and follow-up observations with the \textit{Hubble Space Telescope} (HST) \citep[][]{Romanowsky_24,Tang_25}.  
These studies have highlighted striking similarities between FCC~224 and the DM-deficient galaxies in the NGC~1052 group, suggesting elements of a common origin.
FCC~224 meets the definition of a UDG \citep{vanDokkum_15} within the uncertainties, with an effective radius of $R_{\rm e} = 1.89 \pm 0.01$ kpc, and a central surface brightness of $\mu_{g,0} = 23.97 \pm 0.03$ mag arcsec$^{-2}$. It has a stellar mass of $\log(M_{\star}/M_{\odot}) = 8.24 \pm 0.04$ and is estimated to host $13 \pm 1$ GCs \citep{Tang_25}.

Given these peculiar features, FCC 224 could be a new example of a DM-free dwarf galaxy. This study employs high-resolution Keck Cosmic Web Imager (KCWI) spectroscopy to directly determine the DM content of FCC~224 and investigate its similarities to DF2 and DF4. 
By analysing their kinematics, stellar populations and GC system, we explore the possibility that these galaxies form a new class, and if so, their common traits could serve as a diagnostic framework for identifying additional DM-deficient dwarf galaxies. 

\section{Data and methods}

The Keck/KCWI data for this study were obtained on the night of 14 September 2023 (PI: Forbes), under clear conditions, with seeing ranging from $1.0^{\prime\prime}$ to $1.2^{\prime\prime}$. KCWI was configured with the medium slicer, providing a field of view (FoV) of $16.5^{\prime\prime} \times 20.4^{\prime\prime}$, allowing simultaneous coverage of both the stellar body and six globular clusters (GCs) in FCC 224. The BL grating was centred at 4550\,\AA, covering 3630–5650\,\AA\ in the blue arm, while the RH3 grating was centred at 8600\,\AA, spanning 8205–8962\,\AA\ in the red arm.

Seven blue-arm exposures of 1320 seconds each were obtained with on-chip sky subtraction, while red-arm exposures were shortened to 300 seconds each due to high cosmic ray frequency, resulting in 28 exposures. The total exposure times were $\sim$$2.6$ hours in the blue arm and $\sim$$2.3$ hours in the red arm.
The spectral resolution, measured from arc lamp files, is R = 1908 at 5075\,\AA\ ($\sigma_{\text{inst}} = 66.8$\,km\,s$^{-1}$) for the blue arm and R = 10238 at 8600\,\AA\ ($\sigma_{\text{inst}} = 12.4$\,km\,s$^{-1}$) for the red arm. A full characterisation of the spectral resolution across the full wavelength range is provided in Appendix \ref{appendix:spectral_res}.

For flux calibration of the red-arm data, we used the standard star Feige 110, observed on 12 November 2023, since the calibration file of the originally selected standard star (Feige 25) lacked sufficient sampling in the target wavelength range. Red-arm flat fields from the same 12 November observing run were also used, as flats from 14 September were saturated. While using calibration files from different nights can introduce uncertainties due to potential changes in the instrument, KCWI  is known to be highly stable over time. Over the two-month gap between observations, no significant variations in the performance of the instrument were reported, ensuring that this approach does not affect the reliability of the results. Any residual discrepancies from temporal differences would manifest as small flux calibration offsets, which are negligible compared to the overall uncertainties in the data.

\subsection{Data reduction and sky subtraction} 
Data reduction followed standard Keck/KCWI pipelines \citep[\texttt{KCWIDRP}, ][]{kcwidrp_23}, though additional steps were necessary for cosmic ray removal and sky subtraction, especially for the red arm due to higher cosmic ray frequency and sky-line contamination. The full sky subtraction process is detailed in Appendix \ref{appendix:sky_sub}. 

All spectra were corrected for Milky Way dust attenuation using the \citet{Fitzpatrick_99} reddening law with $R_V = 3.1$, assuming a total extinction of $A_V = 0.031$ mag for FCC 224, based on the recalibration by \citet{Schlafly_11}.

\begin{figure*}
\centering
\includegraphics[width=\textwidth]{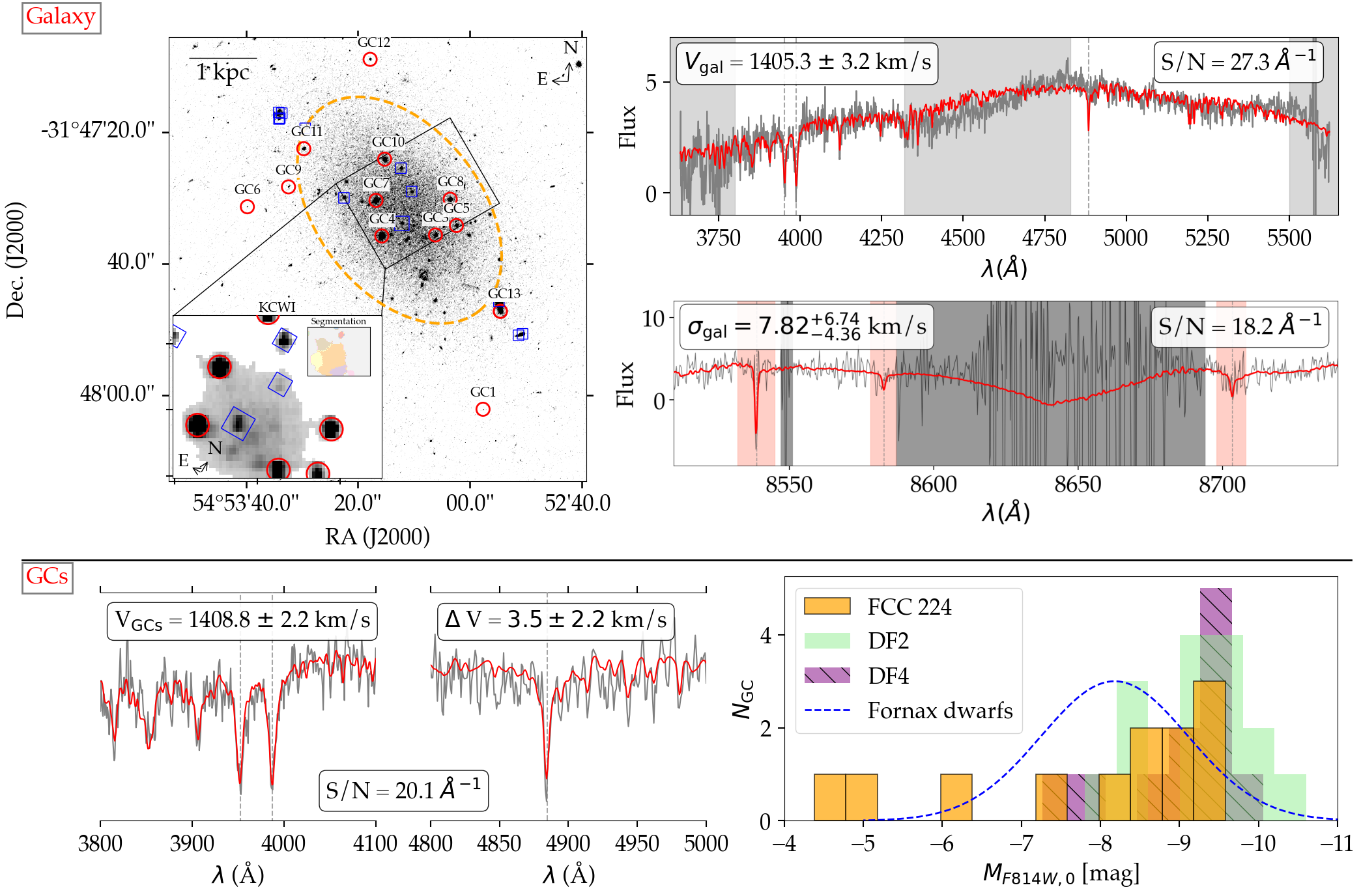}
\caption{Imaging and spectra of FCC 224 (top section) and its GCs (bottom section). \textit{Top left}: HST/WFC3 F814W image,
highlighting the GC candidates in red circles and background/foreground sources in blue squares. The black rectangle and inset axis show the Keck/KCWI  white-light image, along with the segmentation image used to separate galaxy from sky spaxels. The orange dashed ellipse marks the effective radius of the galaxy. \textit{Top right}: Blue and red KCWI spectra of the stellar body of FCC 224, where grey is the observation and red is the \texttt{pPXF} fit. Masked regions of the spectra are marked as grey bands. The CaT lines are marked in the red-arm spectrum by red bands. In both blue- and red-arm spectra, the dashed lines mark key spectral lines (CaH+K and H$\beta$ in the blue-arm spectrum, and the CaT in the red-arm spectrum). \textit{Bottom left and middle:} Close-up views of the CaH+K and H$\beta$ regions of the stacked GC spectrum using the blue arm of KCWI. In all of the panels showing spectra, the S/N is shown, as well as the recovered kinematics (e.g.\ velocity and velocity dispersion) by running \texttt{pPXF}. \textit{Bottom right}:  Comparison of the GCLF of FCC 224 to those of DF2, DF4 and to normal dwarf galaxies in the Fornax cluster as a scaled Gaussian curve \citep{Jordan_15}.}
\label{fig:hst_kcwi}
\end{figure*}

The KCWI FoV covered FCC 224’s central regions, as illustrated in Fig. \ref{fig:hst_kcwi}, where the KCWI pointing is overlaid on the HST/WFC3 F814W imaging. The top right panels display the integrated spectra from both the blue and red arms of KCWI, highlighting key spectral features. 

\subsection{Spectral fitting}

We employed the penalised pixel-fitting (\texttt{pPXF}) code \citep{Cappellari_17} to extract kinematics and stellar populations for FCC 224 and its GCs. \texttt{pPXF} fits observed spectra with combinations of stellar templates convolved with a Gauss-Hermite velocity distribution profile \citep{gerhard}. For the blue-arm data, we used the EMILES stellar population synthesis library \citep{Vazdekis_15}, closely matching the BL grating resolution. For the red arm, we utilised high-resolution templates \citep{Coelho_14} to recover the systemic velocity and velocity dispersion of the galaxy.

To investigate the stellar population gradients, we defined five radial bins centred on the galaxy's core: one inner circular aperture and four additional annuli, each with a width of 5 pixels. These bins were constructed to ensure a minimum signal-to-noise ratio of 15 \AA$^{-1}$ per bin, facilitating reliable measurements. The binning was based on the circularised effective radius ($R_{\rm e, circ} = R_{\rm e}\sqrt{b/a}$, and $b/a = 0.64$) of the galaxy and covers the area within 0.4 $R_{\rm e, circ}$. While it would have been possible to extend the gradient analysis out to the effective radius of the galaxy along the minor axis, this would have required including spaxels previously designated as sky spaxels. To maintain the robustness of our analysis, we limited the study to 0.4 circularised effective radii.

\section{Results}

In this section, we present the results obtained for GC association and GC stellar populations. For the galaxy itself, we report its systemic velocity, stellar populations, age and metallicity gradients, and rotation patterns, with a particular focus on its velocity dispersion to probe the DM content. 

\subsection{Globular clusters}

The KCWI FoV encompasses the six brightest GC candidates around FCC 224, identified by \citet{Tang_25}. We confirmed that they are associated with FCC~224 by measuring radial velocities consistent with those of the galaxy. The spectrum of each confirmed GC and the radial velocity distribution are shown in Appendix \ref{appendix:gc_association}. In the bottom section of Fig \ref{fig:hst_kcwi}, we present the properties of the GCs associated with FCC~224. The left panel shows the blue-arm combined spectra of the six confirmed GCs, while the right panel presents the GCLF of FCC 224 as recovered by \citet{Tang_25}. This GCLF assumes a distance of 20 Mpc \citep{Tang_25}, and reveals an overabundance of luminous GCs, with a top-heavy GCLF similar to those of DF2 and DF4. Additionally, \citet{Tang_25} found that FCC 224’s GCs exhibit a remarkably narrow colour spread of $\sim 0.02$ mag, a property similar to that of the monochromatic GC populations seen in DF2 and DF4 \citep{vanDokkum_22b, Buzzo_23}.

The GC radial distribution in FCC~224 appears consistent with other dwarf galaxies in the Fornax cluster, which typically have $R_{\rm GC}/R_{\rm e} \sim 0.8 $ \citep{Saifollahi_24}, but it is more compact compared to other UDGs \citep[e.g.,][]{Lim_18,Lim_20,Forbes_Gannon_24, Janssens_24}. \citet{Tang_25} reported $R_{\rm GC}/R_{\rm e} = 0.8$, $1.2$ and $2.9$ for FCC~224, DF2 and DF4, respectively. Interestingly, they found evidence of mass segregation in FCC~224, with the most massive GCs located closer to the galaxy centre. This phenomenon could be associated with dynamical friction, similar to what has been observed in other dwarf galaxies \citep[see e.g.,][]{Lotz_01} and even in UDGs \citep{Liang_24}. However, FCC~224 was found to have a shallow surface brightness profile, with a S\'ersic index $n=0.75$ \citep{Tang_25}. Since dynamical friction is expected to have a stronger effect in cuspier profiles, with the DM causing GCs to migrate inwards, and FCC~224 has a shallow profile consistent with a cored potential, then dynamical friction is expected to have a weaker effect. To further explore this hypothesis, we use the simulations of \citet{Doppel_23}, assuming the stellar mass of FCC~224 and a typical GC mass of $\sim 10^5 M_\odot$. These simulations suggest that GC orbital decay is plausible, consistent with suggestions for DF2 by \citet{Nusser_18, Chowdhury_19b, Ogiya_22}, but that the dynamical friction timescale exceeds a Hubble time for most GCs, except those initially located at very small radii ($r < 0.5-1.0$ kpc). For a detailed discussion of GC sizes, radial distributions, mass segregation, and the role of dynamical friction in FCC~224, we refer the reader to \citet{Tang_25}.

After studying the GCs, we masked out all of them to obtain a spectrum of FCC 224’s stellar body. The blue-arm data yielded an integrated spectrum with a signal-to-noise ratio (S/N) of 27.3\,\AA$^{-1}$, while the red-arm spectrum had an S/N of 18\,\AA$^{-1}$.

\begin{figure}
\centering
\includegraphics[width=\columnwidth]{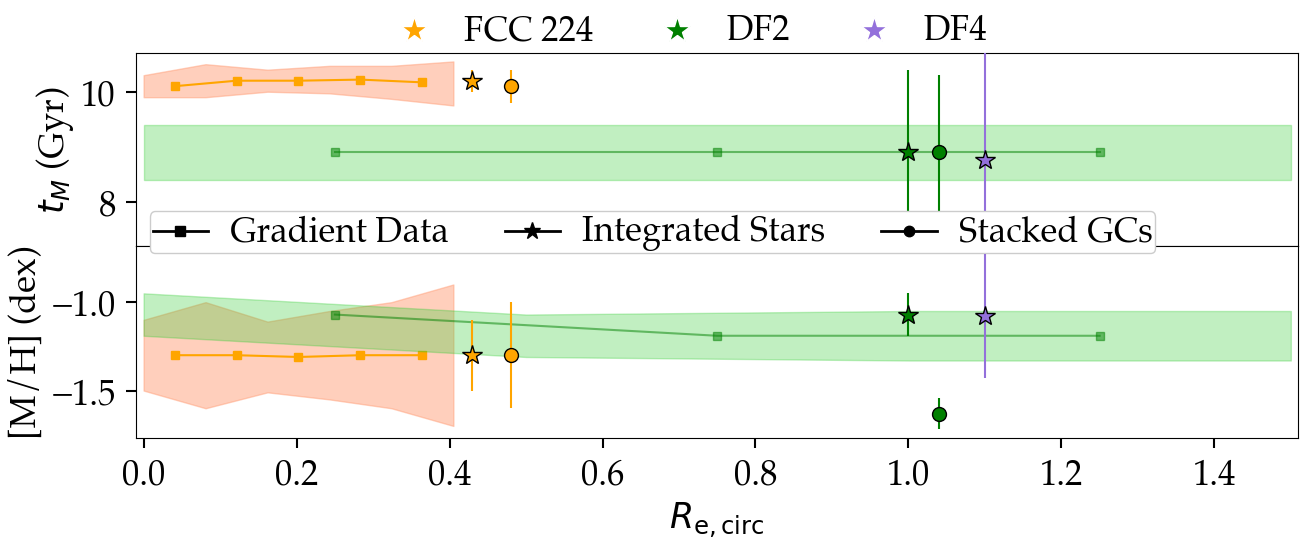}
\caption{Stellar populations of FCC 224 and comparison with DF2 and DF4. Resolved mass-weighted age ($t_M$) and metallicity ([M/H]) profiles for FCC 224 along the circularised effective radius of the galaxy (orange squares, shaded uncertainties) compared with DF2 \citep{Fensch_19} (green squares, shaded). Stars denote integrated stellar populations for FCC 224 (orange), DF2 (green) and DF4 (purple), and circles indicate stacked GC populations for FCC 224 (orange) and DF2 (green). GC stellar population information is not available in the case of DF4. The flat profiles in FCC 224 and DF2 suggest coeval star and GC formation.}
\label{fig:stellar_populations}
\end{figure}

\subsection{Stellar populations} 
We next analyse the kinematics and stellar population properties of FCC~224. This time, we use \texttt{pPXF} coupled with the EMILES libraries \citep{Vazdekis_15} to analyse spectra from the blue arm of KCWI. We measure a systemic velocity for FCC 224 of $1405 \pm 3$~km~s$^{-1}$, excluding regions with high sky contamination or irregular features \citep{Ferre-Mateu_23}. This systemic velocity aligns well with that of the Fornax Cluster \citep[$V = 1425$ km s$^{-1}$, with a velocity dispersion of $\sigma = 374$ km s$^{-1}$,][]{Raj_20,SmithCastelli_24}, indicating that FCC 224 is consistent with being a member of the cluster in velocity space. However, its location at a radial distance of 1.3 Mpc from the cluster centre ($1.8 \times R_{200}$, where $R_{200}$ is the virial radius of the cluster, estimated to be 0.7 Mpc for Fornax \citep{Raj_20,SmithCastelli_24}) suggests that FCC 224 has not yet fallen into the cluster \citep{Rhee_17}. Alternatively, it could be a backsplash galaxy that has passed through the cluster but is now seen on the outskirts \citep{Benavides_21}. 

Disentangling these scenarios is challenging with the current dataset. First-infall galaxies typically exhibit higher star formation rates and greater gas content, whereas backsplash galaxies are often quiescent and gas-depleted due to environmental quenching. However, pre-processing, such as a bullet-dwarf collision, could result in quiescence even for a first-infall galaxy. Additionally, first-infall galaxies usually have higher velocities and lie farther from the cluster centre, while backsplash galaxies decelerate and remain closer. FCC 224, at more than 1.8 virial radii, could favour the first-infall scenario, as backsplash galaxies at such distances are rare \citep{Benavides_21}, though they can extend to 3 virial radii. When looking at stellar populations, backsplash galaxies often show older ages and higher metallicities, whereas FCC 224 exhibits an old stellar population but with very low metallicity (as will be discussed in the next paragraphs), more consistent with pre-processing than with cluster interactions. Ultimately, deeper data (and HI in particular) are required to distinguish between these possibilities.

The blue arm’s broader wavelength range and reduced sky contamination make it ideal for examining both integrated stellar populations, and the radial gradients in annuli of 5 pixels (about 1.3 arcsec), extending out to 0.4 circularised effective radii. Additionally, we stack the spectra of the confirmed GCs, weighting them by flux to assess their stellar populations.

Our results reveal high consistency across FCC 224’s integrated stellar populations, GCs, and radial gradients, with age differences between the galaxy’s bins and GCs remaining below 0.1 Gyr and metallicity differences within 0.05 dex, as shown in Fig. \ref{fig:stellar_populations}.The galaxy's recovered mass-weighted age is $t_M = 10.2$ Gyr, with a mass-weighted metallicity of approximately [M/H] $= -1.3$~dex. These results are equivalent to the ones obtained by \citet{Tang_25} using a combination of photometric data from \textit{HST}, DECaLS and \textit{WISE} data, as well as 30 minutes of KCWI blue arm observations. They found a mass-weighted age of $t_M = 10.1^{+1.3}_{-1.0}$ Gyr, and a metallicity of [M/H] = $-1.25^{+0.05}_{-0.07} $ dex.

\subsection{Kinematics and dynamics}

In this section, we analyse the velocity dispersion of FCC~224 as recovered from both the stars and GC system, and the rotational patterns observed in the galaxy.

\subsubsection{Velocity dispersion from the stars}

Full spectral fitting of the red arm (Fig. \ref{fig:hst_kcwi}), as well as fitting of the three individual Ca II lines using the high-resolution templates from \citet{Coelho_14}, yielded a stellar velocity dispersion of $\sigma_{\rm galaxy}$ = $10.43 \pm 5.76$\,km s$^{-1}$.

The velocity dispersion $\sigma_{\rm galaxy}$ in galaxies like FCC 224 results from both velocity dispersion between member stars and internal motions in the stellar atmospheres: 

\begin{equation}
    \sigma_{\rm galaxy}^2 = \sigma_{\rm stars}^2 + \sigma_{\rm broadening}^2
\end{equation}

In galaxies with substantial DM content, the velocity dispersion tends to be higher due to the gravitational influence of the DM halo. By contrast, DM-deficient galaxies like DF2 and DF4 exhibit notably low velocity dispersions, sometimes reaching levels where intrinsic stellar motions, such as stellar rotation and macroturbulence effects, can noticeably impact the measurements. Estimating intrinsic stellar motions is crucial for FCC 224, given its old, metal-poor stellar population primarily composed of main sequence (MS) and red giant branch (RGB) stars. Since our synthetic models lack intrinsic line broadening, we independently estimated this effect based on empirical studies of low-metallicity stellar kinematics. MS stars in such environments can reach rotation speeds of up to 6 km s$^{-1}$ \citep{Amard_20}, while RGB stars have broadening values between 2 and 11 km s$^{-1}$ \citep{Carney_08, Massarotti_08}. We use an empirical relation between metallicity and broadening \citep{Carney_08} derived for individual metal-poor stars to estimate FCC 224’s internal broadening to be $\sigma_{\rm broadening} = 6.9 \pm 2.6$ km s$^{-1}$, incorporating both rotation and macroturbulence. 

Using the relation from \citet{Carney_08} to interpret the velocity dispersion of FCC 224 comes with several underlying assumptions. First, we extrapolate the rotation and macroturbulence effects observed in individual RGB and MS stars to a composite stellar population. This extrapolation assumes that the stars were formed in a single event, consistent with the minimal metallicity variation observed across the stellar body (see Fig. \ref{fig:stellar_populations}). Additionally, we assume that RGB stars within the galaxy have random orientations, following the approach of \citet{Carney_08}.

Finally, \citet{Carney_08} note that applying their results to integrated light typically requires adjustments to account for metallicity gradients and kinematic variations within a stellar population. However, given the lack of observed metallicity variations across FCC 224 and its classification as a slow rotator, we assume negligible variations across the galaxy's body. Thus, we assume that no adjustments to the Carney relation are necessary in this case.

We subtract this estimated intrinsic broadening from the measured velocity dispersion directly from the spectrum, and find a final stellar velocity dispersion for FCC~224 of $\sigma_{\rm stars} = 7.82^{+6.74}_{-4.36}$ km s$^{-1}$. 

\subsubsection{Velocity dispersion of the GCs} 
The velocity dispersion of the GCs was derived using a Markov Chain Monte Carlo (MCMC) method \citep{Foreman-Mackey_13}. The likelihood function adopted is:

\begin{equation}
L(V_{\rm sys}, \sigma) = \frac{1}{2}\sum_{n=1}^{N_{\rm GC}} \log(2\pi(\sigma^2+\delta v_n^2)) - \sum_{n=1}^{N_{\rm GC}} \frac{(v_n - V_{\rm sys})^2}{2(\sigma^2+\delta v_n^2)},
\end{equation}
where $v_n$ are the GC velocities and $\sigma_{\text{obs}}^2 = \sigma^2 + \delta v^2$, with $\delta v$ as the velocity uncertainty for each GC. 
For our setup, we tested both Jeffreys and uniform priors. The Jeffreys priors have been shown by a number of studies in the literature to result in lower velocity dispersion estimates than the uniform priors \citep{Martin_18, Muller_19, Doppel_21, Toloba_23}. In contrast, the uniform priors were found to overestimate (by $\sim30\%$) the true dispersion in the case of a small number of tracers \citep{Doppel_21,Toloba_23}. 

\begin{figure}
    \centering
    \includegraphics[width=\linewidth]{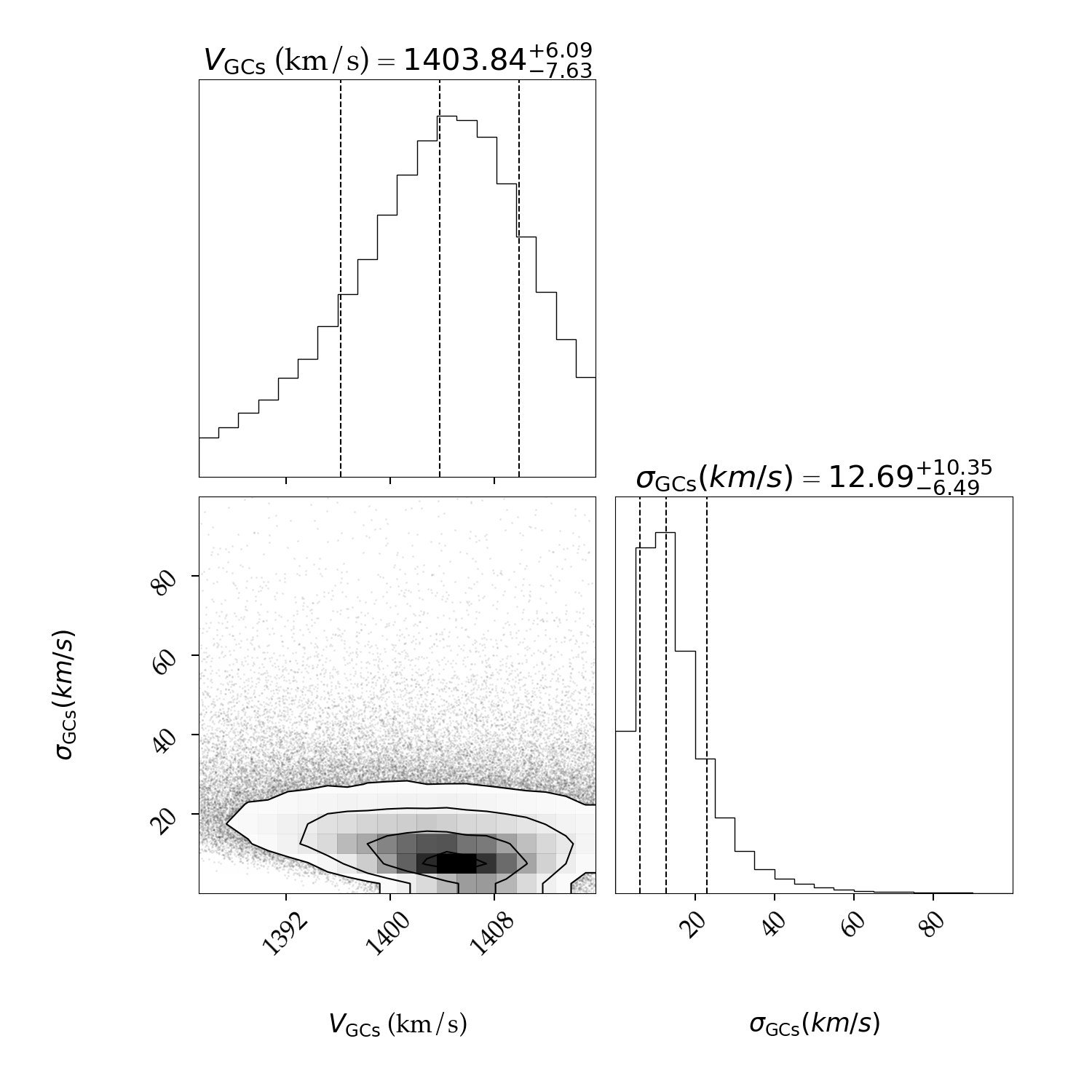}
    \caption{Monte Carlo Markov Chain results of fitting a maximum likelihood function to the velocities recovered for the 5 GCs around FCC 224 with the highest S/N to recover the velocity dispersion and recessional velocity of the host galaxy. The MCMC was run using a uniform prior, and returned $V = 1403.8^{+6.0}_{-7.6}$ km s$^{-1}$ and a dispersion velocity of $\sigma_{\rm GCs} = 12.69^{+10.35}_{-6.49}$ km s$^{-1}$. This velocity dispersion is considered an upper limit based on previous studies that shown that uniform priors tend to overestimate the velocity dispersion of galaxies when a small number of tracers is used.}
    \label{fig:mcmc_velocitydispersion_flat}
\end{figure}

In our case, the Jeffreys priors yielded an unrealistically low velocity dispersion of $0.37^{+1.51}_{-0.31}$ km s$^{-1}$ and a velocity of $V_{\rm sys} = 1407.1^{+3.4}_{-3.5}$ km s$^{-1}$. This extremely low velocity dispersion has been observed in other cases in the literature \citep{Toloba_23}, suggesting that the Jeffreys priors may systematically bias the results toward zero when the true dispersion is small because the priors cannot resolve this extremely low dispersion. 
The uniform priors, on the other hand, yielded a velocity dispersion of $\sigma_{\rm GCs} = 12.69^{+10.35}_{-6.49}$ km. s$^{-1}$ and a velocity of $V_{\rm sys} = 1403.8^{+ 6.1}_{- 7.6}$ km s$^{-1}$. Given that the Jeffreys prior led to an unrealistic velocity dispersion and uniform priors lead to overestimated ones, we chose to adopt the uniform priors result as an upper limit on $\sigma_{\rm GCs}$. The resulting cornerplot of running the MCMC with a uniform prior is shown in Fig. \ref{fig:mcmc_velocitydispersion_flat}.

To validate the accuracy of our methodology, we applied our MCMC routine to the line-of-sight velocities of the GCs around DF2 and DF4, aiming to reproduce the velocity dispersions reported in the literature. For DF2, we obtained $\sigma_{\rm DF2 \, GCs} = 3.7^{+5.3}_{-2.8}$ km s$^{-1}$, and for DF4, we derived $\sigma_{\rm DF4 \, GCs} = 4.1^{+3.7}_{-2.6}$ km s$^{-1}$, both based on uniform priors. These results are consistent, within their uncertainties, with the values previously reported in the literature \citep{vanDokkum_18, vanDokkum_19b}. The slightly lower velocity dispersion for DF2 compared to \citet{vanDokkum_18} remains within the uncertainties and is likely due to random motions and differences in the starting points of the MCMC walkers. 

\begin{figure*}
\centering
\includegraphics[width=\textwidth]{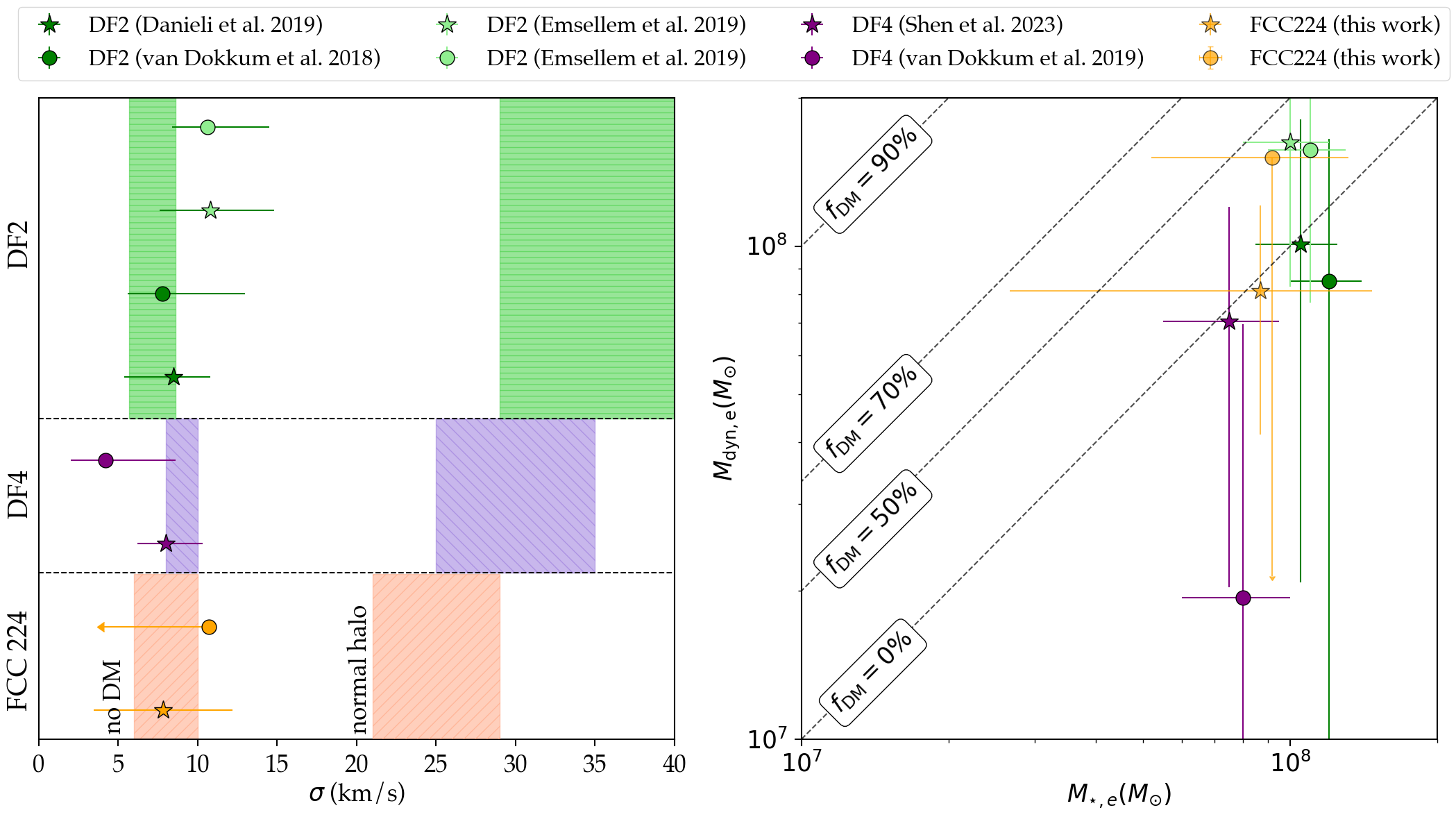}
\caption{Constraints on the velocity dispersion and enclosed mass of FCC 224. \textit{Left:} Stellar velocity dispersion of stars and GCs in FCC 224 (orange), DF2 (green) and DF4 (purple) compared to two models indicated by shades: 1) No DM within the galaxies, and 2) a normal DM halo as predicted by the stellar mass–halo mass relation \citep{Moster_13}. In both panels, the measurements from the stars are shown in star markers and from the GCs in circles \textit{Right:} Comparison of the dynamical and stellar masses within 1 $R_{\rm e}$ for the three DM-deficient galaxies, showing that all three lie near the predicted line for galaxies with a DM fraction close to zero.}
\label{fig:MCMC}
\end{figure*}

In Fig. \ref{fig:MCMC} (left panel), we compare FCC 224’s velocity dispersion with those of DF2 and DF4, using both the GC and stellar measurements. These findings are contextualised within two main frameworks: (1) galaxies with no dynamical evidence for DM, as inferred from their stellar masses, and (2) galaxies that follow the expected stellar mass--halo mass relation from \citet{Moster_13}. 

To obtain the range of predicted velocity dispersions, we follow the prescription thoroughly described in \citet{Wasserman_18}, solving the Jeans equation under the assumption of dynamical equilibrium, an spherical model and isotropic stellar orbits, combining the gravitational potential from both the stellar and dark matter components. To estimate these values, we use the stellar mass-halo mass (SMHM) relation of \citet{Moster_13} to infer that a galaxy with the stellar mass of FCC~224 is well described by a halo mass of $M_{200} = 10^{10.5} M_{\odot}$. For the dark matter halo, we assume an NFW profile \citep{NFW_97}, parametrised by the total halo mass ($M_{200}$) and concentration ($c_{200}$), determined using the mass--concentration relation of \citet{Dutton_Maccio_14}. The stellar density profile was modelled using a S\'ersic distribution, parametrised by the central surface brightness, effective radius, and S\'ersic index, all of the previous were derived by \citet{Tang_25}.
Using these steps, we estimate that the stars alone contribute to the velocity dispersion: $\sigma_{\star} = 7.8 \pm 2.0$ km s$^{-1}$, $7.0^{+1.6}_{-1.3}$ km s$^{-1}$ \citep{Danieli_19} and $9\pm1$ km s$^{-1}$ \citep{Shen_23} for FCC~224, DF2, and DF4, respectively. The predicted velocity dispersions for the galaxies if they follow the SMHM relation are $\sigma_{\rm SMHM} = 25\, \pm \,4$ km s$^{-1}$, $35 \pm 6$ km s$^{-1}$ \citep{Danieli_19} and $30\pm5$ km s$^{-1}$ \citep{Shen_23}. Although this work, \citet{Danieli_19}, and \citet{Shen_23} used different SMHM relations, we verified that our results, based on \citet{Moster_13}, remain consistent within the uncertainties when compared to results obtained using other relations, such as \citet{Behroozi_13}. Additionally, we assumed a cuspy NFW profile for consistency with \citet{Danieli_19} and \citet{Shen_23}. We note, nonetheless, that assuming a cored profile may reduce the tension with the data, which will be explored in a follow-up study using deeper data covering a wider FoV.
Our analysis shows that FCC 224 falls within the `no DM' region, identifying it, along with DF2 and DF4, as a third example of a galaxy with minimal or absent DM content within 1 $R_{\rm e}$.

To further explore FCC 224’s DM content, we convert the velocity dispersions into dynamical masses and compare these with the stellar masses of the galaxies within 1~$R_{\rm e}$. 

According to \cite{Wolf_10}, the dynamical mass is a function of the velocity dispersion and the 3D deprojected effective radius ($r_{1/2}$). Using the 2D projected half-light radius, we have that $r_{1/2} = 4/3 \,\, R_{\rm e}$. Therefore, the equation takes the form:


\begin{equation}
\label{eq:wolf}
M(<r_{1/2}) = 930 \left(\frac{\sigma_{\text{los}}^2}{\text{km}^2 \text{s}^{-2}} \frac{R_e}{\text{pc}}\right) M_\odot,
\end{equation}

where $\sigma_{\rm los}$ is the velocity dispersion along the line-of-sight. Assuming a distance of 20 Mpc to FCC 224 and its effective radius of $R_e = 1.9$\,kpc, we infer an enclosed dynamical mass of $\log(M_{\rm dyn}/M_\odot) = 7.9 \pm 0.4$ from the stars and $\log(M_{\rm dyn}/M_\odot) = 7.5 \pm 0.7$ from the GCs. With a stellar mass of $\log(M_{\star}/M_{\odot}) = 8.24 \pm 0.04$, i.e., a stellar mass within the effective radius of $\log(M_{\star}/M_{\odot}) = 7.94 \pm 0.04$, these results suggest that FCC 224 contains little or no DM within 1 $R_{\rm e}$. 

This comparison is presented in the rightmost panel in Fig. \ref{fig:MCMC}. All three galaxies lie close to the line predicting a DM fraction near zero. For FCC~224, this becomes a small fraction, $\sim$30\%, if we consider the upper limit on dynamical mass from the upper limit on velocity dispersion. For DF2, a similarly small DM fraction can be derived from the measurements of \citet{Emsellem_19}. Although these measurements imply a DM fraction somewhat higher than zero, i.e., not a total lack of DM, they are still considerably smaller than the fractions found for normal dwarf galaxies, known for being highly DM-dominated \citep{Carignan_Beaulieu_89, Persic_96, Cote_00, Swaters_11}. 

Notably, this finding relies on the 20 Mpc distance estimate for FCC 224. Were FCC 224 located closer, the observed dynamics would align with a dark-matter-dominated inner region, as discussed in debates surrounding DF2's distance-dependent DM profile.
DF2 has been subject to ongoing debates about its distance \citep{vanDokkum_18,Trujillo_19,Shen_21}, as uncertainties in distance measurements directly affect its classification as a DM-deficient galaxy. FCC~224 may suffer from the same debates, as measuring the distance of nearby galaxies, where peculiar velocities can dominate over the Hubble flow, introduces significant uncertainties. In such cases, additional measurements of the distance which do not rely only on peculiar velocities need to be pursued. This was done for DF2 using both surface brightness fluctuations \citep{vanDokkum_18b} and the tip of the red giant branch distance using 40 HST orbits \citep{Shen_21}. For FCC~224, the distance was also estimated using surface brightness fluctuations by \citet{Tang_25}. If these galaxies are closer than current estimates suggest, their dynamical mass measurements would rise, bringing the need for DM back into their dynamics. Similarly, a reduced distance would shift their GCLFs back to the expected peak for typical dwarf galaxies. \citet{Tang_25} suggested that FCC~224 would have to be at a distance of $\sim12.5$ Mpc (more than 2$\sigma$ from its measured distance of $18.6 \pm 2.7$ Mpc) to have its GCLF peaking at $M_I = -8.2$ mag, suggested by other Fornax dwarfs \citep[e.g.,][]{Saifollahi_24}. While these distance uncertainties are significant and require further scrutiny, they do not diminish the importance of identifying additional examples. On the contrary, the discovery of more DM-deficient candidates would provide stronger evidence for a common formation mechanism, independent of distance-related controversies. Indeed, as more examples are identified, the statistical reliability of this formation pathway strengthens, emphasising the importance of building a systematic framework for identifying these galaxies. Such a framework is crucial not only for finding further candidates but also for resolving the distance controversy itself by increasing the sample size and refining the conditions under which these galaxies form.

\subsubsection{Rotation} 

Finally, the blue arm data were segmented and rotated by multiples of 30 degrees across the galaxy to identify the axis of maximum rotation. The maximum rotation velocity was found to be $7.5 \pm 3.0$ km s$^{-1}$, which is comparable to the velocity dispersion of the diffuse stellar body. This indicates that the observed rotation is barely discernible from random motions, aligning with the definition of a `slow rotator' as a galaxy lacking clear signs of large-scale rotation \citep{Emsellem_11,Krajnovic_20,Wang_23}. Despite this low rotational velocity, the direction of maximum rotation was found to be orthogonal to the major axis of the galaxy, consistent with prolate rotation given the ellipticity of FCC 224 ($\epsilon = 1-b/a = 0.36$).

\section{Discussion}

\begin{figure}
\centering
\includegraphics[width=\columnwidth]{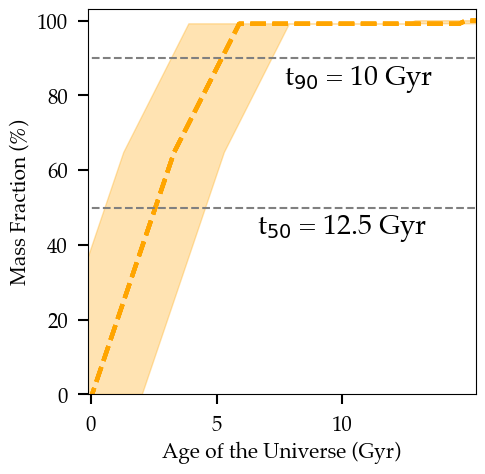}
\caption{Star formation history of FCC 224 indicating early and fast quenching, with 90\% of its stellar mass having formed only $\sim 4.5$ Gyr after the Big Bang (i.e., $t_{90} \sim 10$ Gyr) and a quenching timescale ($t_{50} - t_{90}$) of only 2.5 Gyr. }
\label{fig:SFH}
\end{figure}

\begin{figure*}
    \centering
    \includegraphics[width=\linewidth, trim= 0 7cm 0 0, clip]{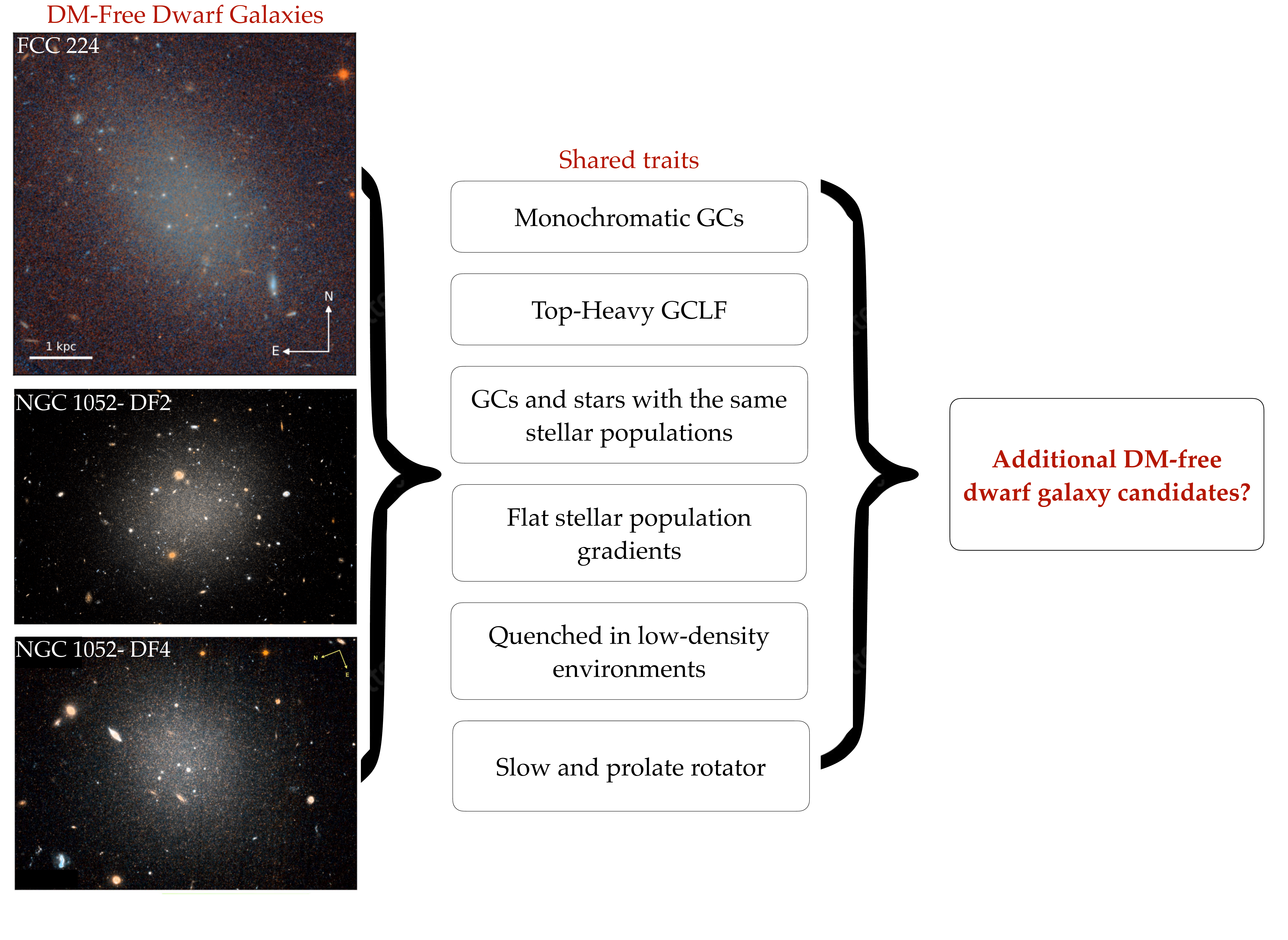}
    \caption{The identified signatures of DM absence in dwarf galaxies. \textit{Left column}: Galaxies found to be deficient of DM within 1 $R_{\rm e}$. \textit{Middle column}: Shared properties of these galaxies that are unusual and different from other dwarf galaxies. \textit{Right column:} We propose that these properties can be used to select DM-free dwarf galaxy candidates from a dedicated survey.}
    \label{fig:diagram_DMabsence}
\end{figure*}

Bringing together all these findings, we see that FCC 224 exhibits several distinctive traits aligning with those of DF2 and DF4, suggesting that these galaxies represent a new, previously unrecognised class, and suggest a `recipe' for identifying other DM-deficient dwarf galaxies.



FCC 224, DF2 and DF4 share a distinctive conformity in the ages and metallicities of their stars and GCs, suggesting coeval formation through an early, distinct star formation event. The uniformity is unusual for dwarf galaxies, where GCs and stars normally exhibit significant age and metallicity differences due to multiple star formation episodes. The finding of constant age and metallicity profiles is also rare among dwarfs, which tend to show older, more metal-poor stars in their outer regions \citep{Koleva_11, Sybilska_17, Bidaran_22, Cardona-Barrero_23, Benavides_24}. Recent studies of UDGs, however, have noted flat or even positive gradients in metallicity \cite[][]{Villaume_22, Ferre-Mateu_25}, similar to FCC 224 and DF2. However, it is worth noting that most observed gradient changes tend to emerge in observations beyond 0.7 $R_{\rm e}$ \citep{Ferre-Mateu_25} and beyond 1 $R_{\rm e}$ in simulations \citep{Benavides_24}. Our findings, based on observations out to 0.4 $R_{\rm e, circ}$, are extrapolated to the outskirts of the galaxy due to the minimal colour variations (F606W$-$F814W) detected, with a difference of less than 0.01 mag across the galaxy. Nevertheless, spectroscopic observations extending further into the galaxy's outskirts will be essential to conclusively confirm the flatness of these stellar population gradients.

From the recovered star formation history (SFH) of FCC~224 (Fig. \ref{fig:SFH}), we conclude that the galaxy is quiescent, with no recent star formation. In addition, we construct the large-scale structure maps around FCC~224, DF2 and DF4 using the 2MASS Extended Source Catalogue \citep[XSC,][]{Huchra_12} and we see that all three of them are particularly far from any massive galaxies or high-density regions, i.e., they are all in low-density environments. Quiescence is particularly rare in low-density environments \citep[$<0.06\%$,][]{Geha_12}, where dwarf galaxies with $M_{\star} \sim 10^8 - 10^9 M_{\odot}$ are predominantly star-forming \citep{Prole_19, Jiaxuan_23, Mao_24, Geha_24}. While a growing number of isolated quiescent galaxies have been identified in recent studies \citep[e.g.,][]{Geha_24,KadoFong_24}, they remain the minority, underscoring the unusual nature of FCC 224. A more detailed characterisation of the environment that the galaxies inhabit and the rarity of this finding is provided in Appendix \ref{appendix_environment}. 

%
To further cement the quiescent state of the galaxy, we note that FCC~224 has no gas (based on analysis of shallow HI Wallaby data \citep{Serra_24} and on the lack of emission lines on its spectrum). DF2 was also shown to be completely devoid of ionised \citep{Fensch_19} and neutral gas \citep{Chowdhury_19a}. Emission lines were also not reported in the spectrum of DF4 \citep{vanDokkum_19b, Shen_23}.

The presence of prolate rotation in FCC~224, a feature also observed in DF2 \citep{Emsellem_19, Lewis_20}, highlights another distinctive characteristic that sets these galaxies apart from typical dwarf galaxies \citep{Lipka_24, Cardona-Barrero_21}. While dwarfs with similar luminosities and ellipticities, particularly dwarf ellipticals in the outskirts of clusters, are often fast, oblate rotators \citep{Lipka_24}, FCC 224 and DF2 exhibit slow, prolate rotation. A similar analysis cannot be done for DF4 due to the lack of spectroscopic measurements, and thus its rotational properties remain uncharacterised. A model of the dynamics of the GC system of DF4, however, suggested no signs of rotation, which was extrapolated to suggest slow stellar rotation in DF4 itself \citep{Yuan_24}. 
Statistically significant prolate rotation has been detected in very few dwarf galaxies to date \citep{Battaglia_22}: And II \citep{Ho_12}, Phoenix \citep{Kacharov_17}, Ursa Minor \citep{Pace_20} and in NGC~6822 \citep{Belland_20}. 
For example, simulations of the rotation patterns of 27 dwarf galaxies using the NIHAO simulations have reported only one to be a prolate rotator, demonstrating how rare this feature is in galaxies in this stellar mass range \citep{Cardona-Barrero_21}. Prolate rotation is often accompanied by a prolate shape \citep{Valenzuela_24} and slow rotation \citep[][]{Schulze_18}, thus, consistent with our findings.


While standard galaxy formation models within the $\Lambda$CDM paradigm allow for DM-deficient galaxies under certain conditions, they typically appear as tidal dwarf galaxies---ejected clumps of baryonic matter resulting from interactions between more massive galaxies \citep{Duc_00,Duc_14, Duc_15, Lelli_15, Ploeckinger_18}. These tidal dwarfs usually host young, metal-rich stellar populations, reflecting the enriched material from which they formed. They also tend to remain close to their progenitors and, in most cases, fall back into them within a short period. For a tidal dwarf galaxy to escape its progenitors’ vicinity, it would require a significantly strong external potential, such as that found in a cluster environment \citep{Ivleva_24}.

As previously discussed, FCC 224, DF2 and DF4 are not consistent with the tidal dwarf scenario \citep{vanDokkum_22, Buzzo_22b, Buzzo_23, Gannon_23b}. Their stellar populations are old, metal-poor, and quenched \citep{vanDokkum_18, Fensch_19, Buzzo_22b}. 
More importantly, the unusual GC systems of these galaxies are inconsistent with tidal dwarf formation scenarios, i.e., no simulation of tidal dwarfs to date could reproduce these GC populations \citep{Haslbauer_19,Mueller_19b,Fensch_19,Fensch_19b}. Tidal stripping of the dwarf’s dark matter component by a massive companion or a cluster \citep{Ogiya_18, Ogiya_22, Maccio_21, Jackson_21, Moreno_22, Ivleva_24} was also proposed as a plausible scenario to explain these galaxies, but similarly, they fail in reproducing the unusual GC populations of the dwarfs.

While a detailed investigation of formation scenarios for these DM-free galaxies is beyond the scope of this paper, our findings provide an opportunity to consider certain scenarios that align better with these observed characteristics. For instance, prolate rotation, a rare feature in dwarf galaxies, is frequently associated with mergers \citep[e.g.][]{Bois_11, Tsatsi_17, Ebrova_17, Hedge_22}. This association raises the possibility that FCC~224, DF2, and DF4 might have formed through such a process. 

The `bullet dwarf' models \citep{Silk_19,vanDokkum_22,Lee_21,Shin_20,Otaki_23,Lee_24} are the only ones in the literature currently relying on mergers (specifically, head-on collisions) as the formation pathway for DM-free dwarfs. 
This scenario predicts the formation of multiple DM-free galaxies simultaneously, as observed in the NGC~1052 group, where DF2 and DF4 are aligned along a trail consistent with this model \citep{Tang_24}. If FCC~224 formed through a similar process, one might expect other galaxies with shared properties to exist nearby.
Indeed, we identify FCC~240, located approximately 13 arcmin from FCC~224, as a compelling candidate for a companion galaxy formed through the same event. FCC~240 shares several striking similarities with FCC~224: it has a comparable ellipticity and position angle, is confirmed to lie at the same distance, and exhibits stellar populations consistent with those of FCC~224 \citep{Tang_25}. Although we lack data on FCC~240’s rotation patterns or velocity dispersion, its seemingly overluminous GC system, stellar populations, and major axis alignment with FCC~224 strongly suggest a shared formation history. These similarities collectively point to the bullet dwarf scenario as the most plausible explanation for the formation of these galaxies.

The model, however, is not entirely capable of explaining the rotation observed in the resulting galaxies. 
The bullet dwarf scenario, being a special case of a head-on collision with a small impact parameter, naturally results in low angular momentum and slow rotation, with any residual rotation likely originating from the progenitor galaxy. While this aligns with the observed slow rotation, it leaves questions about the prolate rotation pattern. Recent studies of prolate rotation in massive galaxies suggest that mergers can disrupt regular rotational patterns depending on the collision angle, producing prolate rotation that can last for many Gyr \citep[][Remus et al. in prep.]{Valenzuela_Remus_24}. It remains to be determined whether this behaviour extends to the mass range of dwarf galaxies. Nevertheless, this provides a compelling conceptual basis for the observed prolate rotation in FCC~224 and DF2, suggesting it could originate from one of the progenitor galaxies having significant initial rotation, which was disrupted during the high-speed collision. Current simulations of the `bullet dwarf' scenario, nonetheless, assume zero initial rotation for the progenitors \citep{Lee_24}. Further refinements in simulations, therefore, are necessary to rigorously test this scenario and to determine whether such conditions can consistently reproduce this newly observed feature of DM-free dwarfs.

Currently, no single formation model fully explains all the observed properties of these galaxies, such as the overluminous GC populations and the slow and prolate rotation, emphasising the need for refined simulations and theoretical models. Although each of these properties are not restricted to DM-deficient galaxies, identifying these shared traits enables a systematic approach for finding DM-free dwarf galaxies, crucial to revealing the processes that allow dwarf galaxies to form and evolve with little or no DM. Fig. \ref{fig:diagram_DMabsence} illustrates the characteristics of the three DM-free galaxies known so far --- the defining traits of this new class of DM-free dwarf galaxies.

\section{Conclusions}
This study investigates FCC 224, a quiescent UDG on the outskirts of the Fornax Cluster, to confirm its DM content and establish its similarities to previously identified DM-deficient galaxies, DF2 and DF4. 

Using high-resolution spectroscopy from the Keck/KCWI and imaging from the \textit{Hubble Space Telescope}, the analysis revealed that FCC 224 is a DM-deficient galaxy, with a stellar velocity dispersion of $\sigma_{\rm stars} = 7.82^{+6.74}_{-4.36}$ km s$^{-1}$, significantly lower than the $25$ km s$^{-1}$ predicted for a typical dwarf galaxy of its stellar mass. The dynamical mass within one effective radius, estimated at $\log(M_{\rm dyn}/M_{\odot}) = 7.9 \pm 0.4$, is consistent with its stellar mass within 1$R_{\rm e}$ of $\log(M_{\star}/M_{\odot}) = 7.94 \pm 0.04$, implying a negligible DM fraction within this region. 

FCC 224's stellar population is uniformly old and metal-poor (i.e., flat age and metallicity profiles), with a mass-weighted age of $10.2 \pm 0.5$ Gyr and metallicity [M/H]$ = -1.3 \pm 0.3$ dex, closely matching those of its GCs. Furthermore, the galaxy hosts an unusual GC system, including a top-heavy GC luminosity function and monochromatic GCs, similar to DF2 and DF4. FCC 224 rotates slowly ($7.5 \pm 3.0$ km s$^{-1}$) and displays prolate rotation, and is quiescent despite residing in a low-density environment, as confirmed by the absence of gas and emission lines. 

These findings align FCC 224 with DF2 and DF4 and signal the existence of a new class of DM-deficient dwarf galaxies, suggesting a broader cosmological relevance that challenges galaxy formation models with the $\Lambda$CDM paradigm and providing a framework to identify additional DM-free galaxies.

\begin{acknowledgements}
We deeply thank the referee for their insightful comments and suggestions which greatly improved the manuscript. We thank Laura Sales for important discussions about dynamical friction. This research was supported by the Australian Research Council Centre of Excellence for All Sky Astrophysics in 3 Dimensions (ASTRO 3D), through project number CE170100013. D.F. and J.B. thank the ARC for support via DP220101863. A.J.R was supported by National Science Foundation grant AST-2308390. AFM has received support from RYC2021-031099-I and PID2021-123313NA-I00 of MICIN/AEI/10.13039/501100011033/FEDER,UE, NextGenerationEU/PRT.
\end{acknowledgements}

\bibliographystyle{aa} 
\bibliography{bibli.bib}

\begin{thebibliography}{122}
\expandafter\ifx\csname natexlab\endcsname\relax\def\natexlab#1{#1}\fi

\bibitem[{{Amard} \& {Matt}(2020)}]{Amard_20}
{Amard}, L. \& {Matt}, S.~P. 2020, \apj, 889, 108

\bibitem[{{Battaglia} \& {Nipoti}(2022)}]{Battaglia_22}
{Battaglia}, G. \& {Nipoti}, C. 2022, Nature Astronomy, 6, 659

\bibitem[{{Behroozi} {et~al.}(2013){Behroozi}, {Wechsler}, \& {Conroy}}]{Behroozi_13}
{Behroozi}, P.~S., {Wechsler}, R.~H., \& {Conroy}, C. 2013, \apj, 770, 57

\bibitem[{{Belland} {et~al.}(2020){Belland}, {Kirby}, {Boylan-Kolchin}, \& {Wheeler}}]{Belland_20}
{Belland}, B., {Kirby}, E., {Boylan-Kolchin}, M., \& {Wheeler}, C. 2020, \apj, 903, 10

\bibitem[{{Benavides} {et~al.}(2021){Benavides}, {Sales}, {Abadi}, {Pillepich}, {Nelson}, {Marinacci}, {Cooper}, {Pakmor}, {Torrey}, {Vogelsberger}, \& {Hernquist}}]{Benavides_21}
{Benavides}, J.~A., {Sales}, L.~V., {Abadi}, M.~G., {et~al.} 2021, Nature Astronomy, 5, 1255

\bibitem[{{Benavides} {et~al.}(2024){Benavides}, {Sales}, {Abadi}, {Vogelsberger}, {Marinacci}, \& {Hernquist}}]{Benavides_24}
{Benavides}, J.~A., {Sales}, L.~V., {Abadi}, M.~G., {et~al.} 2024, \apj, 977, 169

\bibitem[{{Bertin}(2002)}]{Bertin_02}
{Bertin}, E. 2002, in Astronomical Society of the Pacific Conference Series, Vol. 281, Astronomical Data Analysis Software and Systems XI, 228

\bibitem[{{Bidaran} {et~al.}(2022){Bidaran}, {La Barbera}, {Pasquali}, {Peletier}, {van de Ven}, {Grebel}, {Falc{\'o}n-Barroso}, {Sybilska}, {Gadotti}, \& {Coccato}}]{Bidaran_22}
{Bidaran}, B., {La Barbera}, F., {Pasquali}, A., {et~al.} 2022, \mnras, 515, 4622

\bibitem[{{Bois} {et~al.}(2011){Bois}, {Emsellem}, {Bournaud}, {Alatalo}, {Blitz}, {Bureau}, {Cappellari}, {Davies}, {Davis}, {de Zeeuw}, {Duc}, {Khochfar}, {Krajnovi{\'c}}, {Kuntschner}, {Lablanche}, {McDermid}, {Morganti}, {Naab}, {Oosterloo}, {Sarzi}, {Scott}, {Serra}, {Weijmans}, \& {Young}}]{Bois_11}
{Bois}, M., {Emsellem}, E., {Bournaud}, F., {et~al.} 2011, \mnras, 416, 1654

\bibitem[{{Buzzo} {et~al.}(2023){Buzzo}, {Forbes}, {Brodie}, {Janssens}, {Couch}, {Romanowsky}, \& {Gannon}}]{Buzzo_23}
{Buzzo}, M.~L., {Forbes}, D.~A., {Brodie}, J.~P., {et~al.} 2023, \mnras, 522, 595

\bibitem[{{Buzzo} {et~al.}(2022){Buzzo}, {Forbes}, {Brodie}, {Romanowsky}, {Cluver}, {Jarrett}, {Laine}, {Couch}, {Gannon}, {Ferr{\'e}-Mateu}, \& {Okabe}}]{Buzzo_22b}
{Buzzo}, M.~L., {Forbes}, D.~A., {Brodie}, J.~P., {et~al.} 2022, \mnras, 517, 2231

\bibitem[{{Cappellari}(2017)}]{Cappellari_17}
{Cappellari}, M. 2017, \mnras, 466, 798

\bibitem[{{Cardona-Barrero} {et~al.}(2021){Cardona-Barrero}, {Battaglia}, {Di Cintio}, {Revaz}, \& {Jablonka}}]{Cardona-Barrero_21}
{Cardona-Barrero}, S., {Battaglia}, G., {Di Cintio}, A., {Revaz}, Y., \& {Jablonka}, P. 2021, \mnras, 505, L100

\bibitem[{{Cardona-Barrero} {et~al.}(2023){Cardona-Barrero}, {Di Cintio}, {Battaglia}, {Macci{\`o}}, \& {Taibi}}]{Cardona-Barrero_23}
{Cardona-Barrero}, S., {Di Cintio}, A., {Battaglia}, G., {Macci{\`o}}, A.~V., \& {Taibi}, S. 2023, \mnras, 519, 1545

\bibitem[{{Carignan} \& {Beaulieu}(1989)}]{Carignan_Beaulieu_89}
{Carignan}, C. \& {Beaulieu}, S. 1989, \apj, 347, 760

\bibitem[{{Carney} {et~al.}(2008){Carney}, {Gray}, {Yong}, {Latham}, {Manset}, {Zelman}, \& {Laird}}]{Carney_08}
{Carney}, B.~W., {Gray}, D.~F., {Yong}, D., {et~al.} 2008, \aj, 135, 892

\bibitem[{{Chowdhury}(2019)}]{Chowdhury_19a}
{Chowdhury}, A. 2019, \mnras, 482, L99

\bibitem[{{Coelho}(2014)}]{Coelho_14}
{Coelho}, P.~R.~T. 2014, \mnras, 440, 1027

\bibitem[{{C{\^o}t{\'e}} {et~al.}(2000){C{\^o}t{\'e}}, {Carignan}, \& {Freeman}}]{Cote_00}
{C{\^o}t{\'e}}, S., {Carignan}, C., \& {Freeman}, K.~C. 2000, \aj, 120, 3027

\bibitem[{{Danieli} {et~al.}(2019){Danieli}, {van Dokkum}, {Conroy}, {Abraham}, \& {Romanowsky}}]{Danieli_19}
{Danieli}, S., {van Dokkum}, P., {Conroy}, C., {Abraham}, R., \& {Romanowsky}, A.~J. 2019, \apjl, 874, L12

\bibitem[{{Doppel} {et~al.}(2021){Doppel}, {Sales}, {Navarro}, {Abadi}, {Peng}, {Toloba}, \& {Ramos-Almendares}}]{Doppel_21}
{Doppel}, J.~E., {Sales}, L.~V., {Navarro}, J.~F., {et~al.} 2021, \mnras, 502, 1661

\bibitem[{{Doppel} {et~al.}(2023){Doppel}, {Sales}, {Nelson}, {Pillepich}, {Abadi}, {Peng}, {Marinacci}, {Naiman}, {Torrey}, {Vogelsberger}, {Weinberger}, \& {Hernquist}}]{Doppel_23}
{Doppel}, J.~E., {Sales}, L.~V., {Nelson}, D., {et~al.} 2023, \mnras, 518, 2453

\bibitem[{{Duc} {et~al.}(2000){Duc}, {Brinks}, {Springel}, {Pichardo}, {Weilbacher}, \& {Mirabel}}]{Duc_00}
{Duc}, P.~A., {Brinks}, E., {Springel}, V., {et~al.} 2000, \aj, 120, 1238

\bibitem[{{Duc} {et~al.}(2015){Duc}, {Cuillandre}, {Karabal}, {Cappellari}, {Alatalo}, {Blitz}, {Bournaud}, {Bureau}, {Crocker}, {Davies}, {Davis}, {de Zeeuw}, {Emsellem}, {Khochfar}, {Krajnovi{\'c}}, {Kuntschner}, {McDermid}, {Michel-Dansac}, {Morganti}, {Naab}, {Oosterloo}, {Paudel}, {Sarzi}, {Scott}, {Serra}, {Weijmans}, \& {Young}}]{Duc_15}
{Duc}, P.-A., {Cuillandre}, J.-C., {Karabal}, E., {et~al.} 2015, \mnras, 446, 120

\bibitem[{{Duc} {et~al.}(2014){Duc}, {Paudel}, {McDermid}, {Cuillandre}, {Serra}, {Bournaud}, {Cappellari}, \& {Emsellem}}]{Duc_14}
{Duc}, P.-A., {Paudel}, S., {McDermid}, R.~M., {et~al.} 2014, \mnras, 440, 1458

\bibitem[{{Dutta Chowdhury} {et~al.}(2019){Dutta Chowdhury}, {van den Bosch}, \& {van Dokkum}}]{Chowdhury_19b}
{Dutta Chowdhury}, D., {van den Bosch}, F.~C., \& {van Dokkum}, P. 2019, \apj, 877, 133

\bibitem[{{Dutton} \& {Macci{\`o}}(2014)}]{Dutton_Maccio_14}
{Dutton}, A.~A. \& {Macci{\`o}}, A.~V. 2014, \mnras, 441, 3359

\bibitem[{{Ebrov{\'a}} \& {{\L}okas}(2017)}]{Ebrova_17}
{Ebrov{\'a}}, I. \& {{\L}okas}, E.~L. 2017, \apj, 850, 144

\bibitem[{{Eftekhari} {et~al.}(2022){Eftekhari}, {Peletier}, {Scott}, {Mieske}, {Bland-Hawthorn}, {Bryant}, {Cantiello}, {Croom}, {Drinkwater}, {Falc{\'o}n-Barroso}, {Hilker}, {Iodice}, {Napolitano}, {Spavone}, {Valentijn}, {van de Ven}, \& {Venhola}}]{Eftekhari_22}
{Eftekhari}, F.~S., {Peletier}, R.~F., {Scott}, N., {et~al.} 2022, \mnras, 517, 4714

\bibitem[{{Emsellem} {et~al.}(2011){Emsellem}, {Cappellari}, {Krajnovi{\'c}}, {Alatalo}, {Blitz}, {Bois}, {Bournaud}, {Bureau}, {Davies}, {Davis}, {de Zeeuw}, {Khochfar}, {Kuntschner}, {Lablanche}, {McDermid}, {Morganti}, {Naab}, {Oosterloo}, {Sarzi}, {Scott}, {Serra}, {van de Ven}, {Weijmans}, \& {Young}}]{Emsellem_11}
{Emsellem}, E., {Cappellari}, M., {Krajnovi{\'c}}, D., {et~al.} 2011, \mnras, 414, 888

\bibitem[{{Emsellem} {et~al.}(2019){Emsellem}, {van der Burg}, {Fensch}, {Je{\v{r}}{\'a}bkov{\'a}}, {Zanella}, {Agnello}, {Hilker}, {M{\"u}ller}, {Rejkuba}, {Duc}, {Durrell}, {Habas}, {Lelli}, {Lim}, {Marleau}, {Peng}, \& {S{\'a}nchez-Janssen}}]{Emsellem_19}
{Emsellem}, E., {van der Burg}, R. F.~J., {Fensch}, J., {et~al.} 2019, \aap, 625, A76

\bibitem[{{Fensch} {et~al.}(2019{\natexlab{a}}){Fensch}, {Duc}, {Boquien}, {Elmegreen}, {Elmegreen}, {Bournaud}, {Brinks}, {de Grijs}, {Lelli}, {Renaud}, \& {Weilbacher}}]{Fensch_19b}
{Fensch}, J., {Duc}, P.-A., {Boquien}, M., {et~al.} 2019{\natexlab{a}}, \aap, 628, A60

\bibitem[{{Fensch} {et~al.}(2019{\natexlab{b}}){Fensch}, {van der Burg}, {Je{\v{r}}{\'a}bkov{\'a}}, {Emsellem}, {Zanella}, {Agnello}, {Hilker}, {M{\"u}ller}, {Rejkuba}, {Duc}, {Durrell}, {Habas}, {Lim}, {Marleau}, {Peng}, \& {S{\'a}nchez Janssen}}]{Fensch_19}
{Fensch}, J., {van der Burg}, R. F.~J., {Je{\v{r}}{\'a}bkov{\'a}}, T., {et~al.} 2019{\natexlab{b}}, \aap, 625, A77

\bibitem[{{Ferguson}(1989)}]{Ferguson_89}
{Ferguson}, H.~C. 1989, \aj, 98, 367

\bibitem[{{Ferr{\'e}-Mateu} {et~al.}(2025){Ferr{\'e}-Mateu}, {Gannon}, {Forbes}, {Romanowsky}, {Buzzo}, \& {Brodie}}]{Ferre-Mateu_25}
{Ferr{\'e}-Mateu}, A., {Gannon}, J., {Forbes}, D.~A., {et~al.} 2025, arXiv e-prints, arXiv:2501.04088

\bibitem[{{Ferr{\'e}-Mateu} {et~al.}(2023){Ferr{\'e}-Mateu}, {Gannon}, {Forbes}, {Buzzo}, {Romanowsky}, \& {Brodie}}]{Ferre-Mateu_23}
{Ferr{\'e}-Mateu}, A., {Gannon}, J.~S., {Forbes}, D.~A., {et~al.} 2023, \mnras, 526, 4735

\bibitem[{{Fitzpatrick}(1999)}]{Fitzpatrick_99}
{Fitzpatrick}, E.~L. 1999, \pasp, 111, 63

\bibitem[{{Forbes} \& {Gannon}(2024)}]{Forbes_Gannon_24}
{Forbes}, D.~A. \& {Gannon}, J. 2024, \mnras, 528, 608

\bibitem[{{Foreman-Mackey} {et~al.}(2013){Foreman-Mackey}, {Hogg}, {Lang}, \& {Goodman}}]{Foreman-Mackey_13}
{Foreman-Mackey}, D., {Hogg}, D.~W., {Lang}, D., \& {Goodman}, J. 2013, \pasp, 125, 306

\bibitem[{{Gannon} {et~al.}(2023){Gannon}, {Buzzo}, {Ferr{\'e}-Mateu}, {Forbes}, {Brodie}, \& {Romanowsky}}]{Gannon_23b}
{Gannon}, J.~S., {Buzzo}, M.~L., {Ferr{\'e}-Mateu}, A., {et~al.} 2023, \mnras, 524, 2624

\bibitem[{{Geha} {et~al.}(2012){Geha}, {Blanton}, {Yan}, \& {Tinker}}]{Geha_12}
{Geha}, M., {Blanton}, M.~R., {Yan}, R., \& {Tinker}, J.~L. 2012, \apj, 757, 85

\bibitem[{{Geha} {et~al.}(2024){Geha}, {Mao}, {Wechsler}, {Asali}, {Kado-Fong}, {Kallivayalil}, {Nadler}, {Tollerud}, {Weiner}, {de los Reyes}, {Wang}, \& {Wu}}]{Geha_24}
{Geha}, M., {Mao}, Y.-Y., {Wechsler}, R.~H., {et~al.} 2024, \apj, 976, 118

\bibitem[{{Gerhard}(1993)}]{gerhard}
{Gerhard}, O.~E. 1993, \mnras, 265, 213

\bibitem[{{Haslbauer} {et~al.}(2019{\natexlab{a}}){Haslbauer}, {Banik}, {Kroupa}, \& {Grishunin}}]{Haslbauer_19b}
{Haslbauer}, M., {Banik}, I., {Kroupa}, P., \& {Grishunin}, K. 2019{\natexlab{a}}, \mnras, 489, 2634

\bibitem[{{Haslbauer} {et~al.}(2019{\natexlab{b}}){Haslbauer}, {Dabringhausen}, {Kroupa}, {Javanmardi}, \& {Banik}}]{Haslbauer_19}
{Haslbauer}, M., {Dabringhausen}, J., {Kroupa}, P., {Javanmardi}, B., \& {Banik}, I. 2019{\natexlab{b}}, \aap, 626, A47

\bibitem[{{Hegde} {et~al.}(2022){Hegde}, {Bryan}, \& {Genel}}]{Hedge_22}
{Hegde}, S., {Bryan}, G.~L., \& {Genel}, S. 2022, \apj, 937, 38

\bibitem[{{Ho} {et~al.}(2012){Ho}, {Geha}, {Munoz}, {Guhathakurta}, {Kalirai}, {Gilbert}, {Tollerud}, {Bullock}, {Beaton}, \& {Majewski}}]{Ho_12}
{Ho}, N., {Geha}, M., {Munoz}, R.~R., {et~al.} 2012, \apj, 758, 124

\bibitem[{{Huchra} {et~al.}(2012){Huchra}, {Macri}, {Masters}, {Jarrett}, {Berlind}, {Calkins}, {Crook}, {Cutri}, {Erdo{\v{g}}du}, {Falco}, {George}, {Hutcheson}, {Lahav}, {Mader}, {Mink}, {Martimbeau}, {Schneider}, {Skrutskie}, {Tokarz}, \& {Westover}}]{Huchra_12}
{Huchra}, J.~P., {Macri}, L.~M., {Masters}, K.~L., {et~al.} 2012, \apjs, 199, 26

\bibitem[{{Ivleva} {et~al.}(2024){Ivleva}, {Remus}, {Valenzuela}, \& {Dolag}}]{Ivleva_24}
{Ivleva}, A., {Remus}, R.-S., {Valenzuela}, L.~M., \& {Dolag}, K. 2024, \aap, 687, A105

\bibitem[{{Jackson} {et~al.}(2021){Jackson}, {Kaviraj}, {Martin}, {Devriendt}, {Slyz}, {Silk}, {Dubois}, {Yi}, {Pichon}, {Volonteri}, {Choi}, {Kimm}, {Kraljic}, \& {Peirani}}]{Jackson_21}
{Jackson}, R.~A., {Kaviraj}, S., {Martin}, G., {et~al.} 2021, \mnras, 502, 1785

\bibitem[{{Janssens} {et~al.}(2024){Janssens}, {Forbes}, {Romanowsky}, {Gannon}, {Pfeffer}, {Couch}, {Brodie}, {Harris}, {Durrell}, \& {Bekki}}]{Janssens_24}
{Janssens}, S.~R., {Forbes}, D.~A., {Romanowsky}, A.~J., {et~al.} 2024, \mnras, 534, 783

\bibitem[{{Jord{\'a}n} {et~al.}(2015){Jord{\'a}n}, {Peng}, {Blakeslee}, {C{\^o}t{\'e}}, {Eyheramendy}, \& {Ferrarese}}]{Jordan_15}
{Jord{\'a}n}, A., {Peng}, E.~W., {Blakeslee}, J.~P., {et~al.} 2015, \apjs, 221, 13

\bibitem[{{Kacharov} {et~al.}(2017){Kacharov}, {Battaglia}, {Rejkuba}, {Cole}, {Carrera}, {Fraternali}, {Wilkinson}, {Gallart}, {Irwin}, \& {Tolstoy}}]{Kacharov_17}
{Kacharov}, N., {Battaglia}, G., {Rejkuba}, M., {et~al.} 2017, \mnras, 466, 2006

\bibitem[{{Kado-Fong} {et~al.}(2024){Kado-Fong}, {Robinson}, {Nyland}, {Greene}, {Suess}, {Stierwalt}, \& {Beaton}}]{KadoFong_24}
{Kado-Fong}, E., {Robinson}, A., {Nyland}, K., {et~al.} 2024, \apj, 963, 37

\bibitem[{{Koleva} {et~al.}(2011){Koleva}, {Prugniel}, {De Rijcke}, \& {Zeilinger}}]{Koleva_11}
{Koleva}, M., {Prugniel}, P., {De Rijcke}, S., \& {Zeilinger}, W.~W. 2011, \mnras, 417, 1643

\bibitem[{{Krajnovi{\'c}} {et~al.}(2020){Krajnovi{\'c}}, {Ural}, {Kuntschner}, {Goudfrooij}, {Wolfe}, {Cappellari}, {Davies}, {de Zeeuw}, {Duc}, {Emsellem}, {Karick}, {McDermid}, {Mei}, \& {Naab}}]{Krajnovic_20}
{Krajnovi{\'c}}, D., {Ural}, U., {Kuntschner}, H., {et~al.} 2020, \aap, 635, A129

\bibitem[{{Lee} {et~al.}(2021){Lee}, {Shin}, \& {Kim}}]{Lee_21}
{Lee}, J., {Shin}, E.-j., \& {Kim}, J.-h. 2021, \apjl, 917, L15

\bibitem[{{Lee} {et~al.}(2024){Lee}, {Shin}, {Kim}, {Shapiro}, \& {Chung}}]{Lee_24}
{Lee}, J., {Shin}, E.-j., {Kim}, J.-h., {Shapiro}, P.~R., \& {Chung}, E. 2024, \apj, 966, 72

\bibitem[{{Lelli} {et~al.}(2015){Lelli}, {Duc}, {Brinks}, {Bournaud}, {McGaugh}, {Lisenfeld}, {Weilbacher}, {Boquien}, {Revaz}, {Braine}, {Koribalski}, \& {Belles}}]{Lelli_15}
{Lelli}, F., {Duc}, P.-A., {Brinks}, E., {et~al.} 2015, \aap, 584, A113

\bibitem[{{Lewis} {et~al.}(2020){Lewis}, {Brewer}, \& {Wan}}]{Lewis_20}
{Lewis}, G.~F., {Brewer}, B.~J., \& {Wan}, Z. 2020, \mnras, 491, L1

\bibitem[{{Li} {et~al.}(2023){Li}, {Greene}, {Greco}, {Beaton}, {Danieli}, {Goulding}, {Huang}, \& {Kado-Fong}}]{Jiaxuan_23}
{Li}, J., {Greene}, J.~E., {Greco}, J., {et~al.} 2023, \apj, 955, 2

\bibitem[{{Li} {et~al.}(2024){Li}, {Brewer}, \& {Lewis}}]{Yuan_24}
{Li}, Y., {Brewer}, B.~J., \& {Lewis}, G. 2024, \pasa, 41, e073

\bibitem[{{Liang} {et~al.}(2024){Liang}, {Jiang}, {Danieli}, {Benson}, \& {Hopkins}}]{Liang_24}
{Liang}, J., {Jiang}, F., {Danieli}, S., {Benson}, A., \& {Hopkins}, P. 2024, \apj, 964, 53

\bibitem[{{Lim} {et~al.}(2020){Lim}, {C{\^o}t{\'e}}, {Peng}, {Ferrarese}, {Roediger}, {Durrell}, {Mihos}, {Wang}, {Gwyn}, {Cuillandre}, {Liu}, {S{\'a}nchez-Janssen}, {Toloba}, {Sales}, {Guhathakurta}, {Lan{\c{c}}on}, \& {Puzia}}]{Lim_20}
{Lim}, S., {C{\^o}t{\'e}}, P., {Peng}, E.~W., {et~al.} 2020, \apj, 899, 69

\bibitem[{{Lim} {et~al.}(2018){Lim}, {Peng}, {C{\^o}t{\'e}}, {Sales}, {den Brok}, {Blakeslee}, \& {Guhathakurta}}]{Lim_18}
{Lim}, S., {Peng}, E.~W., {C{\^o}t{\'e}}, P., {et~al.} 2018, \apj, 862, 82

\bibitem[{{Lipka} {et~al.}(2024){Lipka}, {Thomas}, {Saglia}, {Bender}, {Fabricius}, \& {Partmann}}]{Lipka_24}
{Lipka}, M., {Thomas}, J., {Saglia}, R., {et~al.} 2024, \apj, 976, 17

\bibitem[{{Lotz} {et~al.}(2001){Lotz}, {Telford}, {Ferguson}, {Miller}, {Stiavelli}, \& {Mack}}]{Lotz_01}
{Lotz}, J.~M., {Telford}, R., {Ferguson}, H.~C., {et~al.} 2001, \apj, 552, 572

\bibitem[{{Macci{\`o}} {et~al.}(2021){Macci{\`o}}, {Prats}, \& {Dixon}}]{Maccio_21}
{Macci{\`o}}, A.~V., {Prats}, D.~H., \& {Dixon}, K.~L. 2021, \mnras, 501, 693

\bibitem[{{Mao} {et~al.}(2024){Mao}, {Geha}, {Wechsler}, {Asali}, {Wang}, {Kado-Fong}, {Kallivayalil}, {Nadler}, {Tollerud}, {Weiner}, {de los Reyes}, \& {Wu}}]{Mao_24}
{Mao}, Y.-Y., {Geha}, M., {Wechsler}, R.~H., {et~al.} 2024, \apj, 976, 117

\bibitem[{{Martin} {et~al.}(2018){Martin}, {Collins}, {Longeard}, \& {Tollerud}}]{Martin_18}
{Martin}, N.~F., {Collins}, M. L.~M., {Longeard}, N., \& {Tollerud}, E. 2018, \apjl, 859, L5

\bibitem[{{Massarotti} {et~al.}(2008){Massarotti}, {Latham}, {Stefanik}, \& {Fogel}}]{Massarotti_08}
{Massarotti}, A., {Latham}, D.~W., {Stefanik}, R.~P., \& {Fogel}, J. 2008, \aj, 135, 209

\bibitem[{{Moreno} {et~al.}(2022){Moreno}, {Danieli}, {Bullock}, {Feldmann}, {Hopkins}, {{\c{c}}atmabacak}, {Gurvich}, {Lazar}, {Klein}, {Hummels}, {Hafen}, {Mercado}, {Yu}, {Jiang}, {Wheeler}, {Wetzel}, {Angl{\'e}s-Alc{\'a}zar}, {Boylan-Kolchin}, {Quataert}, {Faucher-Gigu{\`e}re}, \& {Kere{\v{s}}}}]{Moreno_22}
{Moreno}, J., {Danieli}, S., {Bullock}, J.~S., {et~al.} 2022, Nature Astronomy, 6, 496

\bibitem[{{Moster} {et~al.}(2013){Moster}, {Naab}, \& {White}}]{Moster_13}
{Moster}, B.~P., {Naab}, T., \& {White}, S. D.~M. 2013, \mnras, 428, 3121

\bibitem[{{M{\"u}ller} {et~al.}(2019{\natexlab{a}}){M{\"u}ller}, {Famaey}, \& {Zhao}}]{Muller_19}
{M{\"u}ller}, O., {Famaey}, B., \& {Zhao}, H. 2019{\natexlab{a}}, \aap, 623, A36

\bibitem[{{M{\"u}ller} {et~al.}(2019{\natexlab{b}}){M{\"u}ller}, {Famaey}, \& {Zhao}}]{Mueller_19b}
{M{\"u}ller}, O., {Famaey}, B., \& {Zhao}, H. 2019{\natexlab{b}}, \aap, 623, A36

\bibitem[{{Navarro} {et~al.}(1997){Navarro}, {Frenk}, \& {White}}]{NFW_97}
{Navarro}, J.~F., {Frenk}, C.~S., \& {White}, S. D.~M. 1997, \apj, 490, 493

\bibitem[{{Neill} {et~al.}(2023){Neill}, {Matuszewski}, {Martin}, {Brodheim}, \& {Rizzi}}]{kcwidrp_23}
{Neill}, D., {Matuszewski}, M., {Martin}, C., {Brodheim}, M., \& {Rizzi}, L. 2023, {KCWI\_DRP: Keck Cosmic Web Imager Data Reduction Pipeline in Python}, Astrophysics Source Code Library, record ascl:2301.019

\bibitem[{{Nusser}(2018)}]{Nusser_18}
{Nusser}, A. 2018, \apjl, 863, L17

\bibitem[{{Ogiya}(2018)}]{Ogiya_18}
{Ogiya}, G. 2018, \mnras, 480, L106

\bibitem[{{Ogiya} {et~al.}(2022){Ogiya}, {van den Bosch}, \& {Burkert}}]{Ogiya_22}
{Ogiya}, G., {van den Bosch}, F.~C., \& {Burkert}, A. 2022, \mnras, 510, 2724

\bibitem[{{Otaki} \& {Mori}(2023)}]{Otaki_23}
{Otaki}, K. \& {Mori}, M. 2023, \mnras, 525, 2535

\bibitem[{{Pace} {et~al.}(2020){Pace}, {Kaplinghat}, {Kirby}, {Simon}, {Tollerud}, {Mu{\~n}oz}, {C{\^o}t{\'e}}, {Djorgovski}, \& {Geha}}]{Pace_20}
{Pace}, A.~B., {Kaplinghat}, M., {Kirby}, E., {et~al.} 2020, \mnras, 495, 3022

\bibitem[{{Persic} {et~al.}(1996){Persic}, {Salucci}, \& {Stel}}]{Persic_96}
{Persic}, M., {Salucci}, P., \& {Stel}, F. 1996, \mnras, 281, 27

\bibitem[{{Ploeckinger} {et~al.}(2018){Ploeckinger}, {Sharma}, {Schaye}, {Crain}, {Schaller}, \& {Barber}}]{Ploeckinger_18}
{Ploeckinger}, S., {Sharma}, K., {Schaye}, J., {et~al.} 2018, \mnras, 474, 580

\bibitem[{{Prole} {et~al.}(2019){Prole}, {van der Burg}, {Hilker}, \& {Davies}}]{Prole_19}
{Prole}, D.~J., {van der Burg}, R.~F.~J., {Hilker}, M., \& {Davies}, J.~I. 2019, \mnras, 488, 2143

\bibitem[{{Raj} {et~al.}(2020){Raj}, {Iodice}, {Napolitano}, {Hilker}, {Spavone}, {Peletier}, {Su}, {Falc{\'o}n-Barroso}, {van de Ven}, {Cantiello}, {Kleiner}, {Venhola}, {Mieske}, {Paolillo}, {Capaccioli}, \& {Schipani}}]{Raj_20}
{Raj}, M.~A., {Iodice}, E., {Napolitano}, N.~R., {et~al.} 2020, \aap, 640, A137

\bibitem[{{Rhee} {et~al.}(2017){Rhee}, {Smith}, {Choi}, {Yi}, {Jaff{\'e}}, {Candlish}, \& {S{\'a}nchez-J{\'a}nssen}}]{Rhee_17}
{Rhee}, J., {Smith}, R., {Choi}, H., {et~al.} 2017, \apj, 843, 128

\bibitem[{{Romanowsky} {et~al.}(2024){Romanowsky}, {Cabrera}, \& {Janssens}}]{Romanowsky_24}
{Romanowsky}, A.~J., {Cabrera}, E., \& {Janssens}, S.~R. 2024, Research Notes of the American Astronomical Society, 8, 202

\bibitem[{{Romero-G{\'o}mez} {et~al.}(2023){Romero-G{\'o}mez}, {Peletier}, {Aguerri}, {Mieske}, {Scott}, {Bland-Hawthorn}, {Bryant}, {Croom}, {Eftekhari}, {Falc{\'o}n-Barroso}, {Hilker}, {van de Ven}, \& {Venhola}}]{RomeroGomez_23}
{Romero-G{\'o}mez}, J., {Peletier}, R.~F., {Aguerri}, J.~A.~L., {et~al.} 2023, \mnras, 522, 130

\bibitem[{{Saifollahi} {et~al.}(2024){Saifollahi}, {Voggel}, {Lan{\c{c}}on}, {Cantiello}, {Raj}, {Cuillandre}, {Larsen}, {Marleau}, {Venhola}, {Schirmer}, {Carollo}, {Duc}, {Ferguson}, {Hunt}, {K{\"u}mmel}, {Laureijs}, {Marchal}, {Nucita}, {Peletier}, {Poulain}, {Rejkuba}, {S{\'a}nchez-Janssen}, {Urbano}, {Abdurro'uf}, {Altieri}, {Baes}, {Bolzonella}, {Conselice}, {Cote}, {Dimauro}, {Gonzalez}, {Habas}, {Hudelot}, {Kluge}, {Lonare}, {Massari}, {Romelli}, {Scaramella}, {Sola}, {Stone}, {Tortora}, {van Mierlo}, {Knapen}, {Mart{\'\i}n-Fleitas}, {Mora}, {Rom{\'a}n}, {Aghanim}, {Amara}, {Andreon}, {Auricchio}, {Baldi}, {Balestra}, {Bardelli}, {Basset}, {Bender}, {Bonino}, {Branchini}, {Brescia}, {Brinchmann}, {Camera}, {Capobianco}, {Carbone}, {Carretero}, {Casas}, {Castellano}, {Cavuoti}, {Cimatti}, {Congedo}, {Conversi}, {Copin}, {Courbin}, {Courtois}, {Cropper}, {Da Silva}, {Degaudenzi}, {Di Giorgio}, {Dinis}, {Dubath}, {Dupac}, {Dusini}, {Fabricius}, {Farina}, {Farrens}, {Ferriol}, {Fosalba}, {Frailis},
  {Franceschi}, {Fumana}, {Galeotta}, {Garilli}, {Gillard}, {Gillis}, {Giocoli}, {G{\'o}mez-Alvarez}, {Granett}, {Grazian}, {Grupp}, {Guzzo}, {Haugan}, {Hoar}, {Hoekstra}, {Holmes}, {Hook}, {Hormuth}, {Hornstrup}, {Jahnke}, {Jhabvala}, {Keih{\"a}nen}, {Kermiche}, {Kiessling}, {Kitching}, {Kohley}, {Kubik}, {Kuijken}, {Kunz}, {Kurki-Suonio}, {Lahav}, {Le Mignant}, {Ligori}, {Lilje}, {Lindholm}, {Lloro}, {Maino}, {Maiorano}, {Mansutti}, {Marggraf}, {Markovic}, {Martinet}, {Marulli}, {Massey}, {Maurogordato}, {McCracken}, {Medinaceli}, {Mei}, {Melchior}, {Mellier}, {Meneghetti}, {Meylan}, {Moresco}, {Moscardini}, {Munari}, {Nakajima}, {Nichol}, {Niemi}, {Padilla}, {Paltani}, {Pasian}, {Pedersen}, {Percival}, {Pettorino}, {Pires}, {Polenta}, {Poncet}, {Popa}, {Pozzetti}, {Racca}, {Raison}, {Rebolo}, {Refregier}, {Renzi}, {Rhodes}, {Riccio}, {Roncarelli}, {Rossetti}, {Saglia}, {Sapone}, {Sartoris}, {Schneider}, {Schrabback}, {Secroun}, {Seidel}, {Serrano}, {Sirignano}, {Sirri}, {Stanco}, {Tallada-Cresp{\'\i}},
  {Taylor}, {Teplitz}, {Tereno}, {Toledo-Moreo}, {Torradeflot}, {Tsyganov}, {Tutusaus}, {Valentijn}, {Valenziano}, {Vassallo}, {Verdoes Kleijn}, {Veropalumbo}, {Wang}, {Weller}, {Williams}, {Zamorani}, {Zucca}, {Biviano}, {Burigana}, {Scottez}, {Simon}, {Balogh}, \& {Scott}}]{Saifollahi_24}
{Saifollahi}, T., {Voggel}, K., {Lan{\c{c}}on}, A., {et~al.} 2024, arXiv e-prints, arXiv:2405.13500

\bibitem[{{Schlafly} \& {Finkbeiner}(2011)}]{Schlafly_11}
{Schlafly}, E.~F. \& {Finkbeiner}, D.~P. 2011, \apj, 737, 103

\bibitem[{{Schulze} {et~al.}(2018){Schulze}, {Remus}, {Dolag}, {Burkert}, {Emsellem}, \& {van de Ven}}]{Schulze_18}
{Schulze}, F., {Remus}, R.-S., {Dolag}, K., {et~al.} 2018, \mnras, 480, 4636

\bibitem[{{Scott} {et~al.}(2020){Scott}, {Eftekhari}, {Peletier}, {Bryant}, {Bland-Hawthorn}, {Capaccioli}, {Croom}, {Drinkwater}, {Falc{\'o}n-Barroso}, {Hilker}, {Iodice}, {Lorente}, {Mieske}, {Spavone}, {van de Ven}, \& {Venhola}}]{Scott_20}
{Scott}, N., {Eftekhari}, F.~S., {Peletier}, R.~F., {et~al.} 2020, \mnras, 497, 1571

\bibitem[{{Serra} {et~al.}(2023){Serra}, {Maccagni}, {Kleiner}, {Moln{\'a}r}, {Ramatsoku}, {Loni}, {Loi}, {de Blok}, {Bryan}, {Dettmar}, {Frank}, {van Gorkom}, {Govoni}, {Iodice}, {J{\'o}zsa}, {Kamphuis}, {Kraan-Korteweg}, {Loubser}, {Murgia}, {Oosterloo}, {Peletier}, {Pisano}, {Smith}, {Trager}, \& {Verheijen}}]{Serra_24}
{Serra}, P., {Maccagni}, F.~M., {Kleiner}, D., {et~al.} 2023, \aap, 673, A146

\bibitem[{{Shen} {et~al.}(2021{\natexlab{a}}){Shen}, {Danieli}, {van Dokkum}, {Abraham}, {Brodie}, {Conroy}, {Dolphin}, {Romanowsky}, {Kruijssen}, \& {Dutta Chowdhury}}]{Shen_21}
{Shen}, Z., {Danieli}, S., {van Dokkum}, P., {et~al.} 2021{\natexlab{a}}, \apjl, 914, L12

\bibitem[{{Shen} {et~al.}(2021{\natexlab{b}}){Shen}, {van Dokkum}, \& {Danieli}}]{Shen_21a}
{Shen}, Z., {van Dokkum}, P., \& {Danieli}, S. 2021{\natexlab{b}}, \apj, 909, 179

\bibitem[{{Shen} {et~al.}(2023){Shen}, {van Dokkum}, \& {Danieli}}]{Shen_23}
{Shen}, Z., {van Dokkum}, P., \& {Danieli}, S. 2023, \apj, 957, 6

\bibitem[{{Shin} {et~al.}(2020){Shin}, {Jung}, {Kwon}, {Kim}, {Lee}, {Jo}, \& {Oh}}]{Shin_20}
{Shin}, E.-j., {Jung}, M., {Kwon}, G., {et~al.} 2020, \apj, 899, 25

\bibitem[{{Silk}(2019)}]{Silk_19}
{Silk}, J. 2019, \mnras, 488, L24

\bibitem[{{Smith Castelli} {et~al.}(2024){Smith Castelli}, {Cortesi}, {Haack}, {Lopes}, {Thain{\'a}-Batista}, {Cid Fernandes}, {Lomel{\'\i}-N{\'u}{\~n}ez}, {Ribeiro}, {de Bom}, {Cernic}, {Sodr{\'e}}, {Zenocratti}, {De Rossi}, {Calder{\'o}n}, {Herpich}, {Telles}, {Saha}, {Lopes}, {Lopes-Silva}, {Gon{\c{c}}alves}, {Bambrila}, {Cardoso}, {Buzzo}, {Astudillo Sotomayor}, {Demarco}, {Leigh}, {Sarzi}, {Men{\'e}ndez-Delmestre}, {Faifer}, {Jim{\'e}nez-Teja}, {Grossi}, {Hern{\'a}ndez-Jim{\'e}nez}, {Krabbe}, {Guti{\'e}rrez Soto}, {Brand{\~a}o}, {Espinosa}, {Olave-Rojas}, {Oliveira Schwarz}, {Almeida-Fernandes}, {Schoenell}, {Ribeiro}, {Kanaan}, \& {Mendes de Oliveira}}]{SmithCastelli_24}
{Smith Castelli}, A.~V., {Cortesi}, A., {Haack}, R.~F., {et~al.} 2024, \mnras, 530, 3787

\bibitem[{{Soto} {et~al.}(2016){Soto}, {Lilly}, {Bacon}, {Richard}, \& {Conseil}}]{Soto_16}
{Soto}, K.~T., {Lilly}, S.~J., {Bacon}, R., {Richard}, J., \& {Conseil}, S. 2016, \mnras, 458, 3210

\bibitem[{{Swaters} {et~al.}(2011){Swaters}, {Sancisi}, {van Albada}, \& {van der Hulst}}]{Swaters_11}
{Swaters}, R.~A., {Sancisi}, R., {van Albada}, T.~S., \& {van der Hulst}, J.~M. 2011, \apj, 729, 118

\bibitem[{{Sybilska} {et~al.}(2017){Sybilska}, {Lisker}, {Kuntschner}, {Vazdekis}, {van de Ven}, {Peletier}, {Falc{\'o}n-Barroso}, {Vijayaraghavan}, \& {Janz}}]{Sybilska_17}
{Sybilska}, A., {Lisker}, T., {Kuntschner}, H., {et~al.} 2017, \mnras, 470, 815

\bibitem[{{Tang} {et~al.}(2025{\natexlab{a}}){Tang}, {Romanowsky}, {Gannon}, {Janssens}, {Brodie}, {Bundy}, {Buzzo}, {Cabrera}, {Danieli}, {Ferr{\'e}-Mateu}, {Forbes}, \& {van Dokkum}}]{Tang_25}
{Tang}, Y., {Romanowsky}, A.~J., {Gannon}, J.~S., {et~al.} 2025{\natexlab{a}}, arXiv e-prints, arXiv:2501.10665

\bibitem[{{Tang} {et~al.}(2025{\natexlab{b}}){Tang}, {Romanowsky}, {van Dokkum}, {Jarrett}, {Bundy}, {Buzzo}, {Danieli}, {Gannon}, {Keim}, {Laine}, \& {Shen}}]{Tang_24}
{Tang}, Y., {Romanowsky}, A.~J., {van Dokkum}, P.~G., {et~al.} 2025{\natexlab{b}}, \apj, 978, 21

\bibitem[{{Tanoglidis} {et~al.}(2021){Tanoglidis}, {Drlica-Wagner}, {Wei}, {Li}, {S{\'a}nchez}, {Zhang}, {Peter}, {Feldmeier-Krause}, {Prat}, {Casey}, {Palmese}, {S{\'a}nchez}, {DeRose}, {Conselice}, {Gagnon}, {Abbott}, {Aguena}, {Allam}, {Avila}, {Bechtol}, {Bertin}, {Bhargava}, {Brooks}, {Burke}, {Rosell}, {Kind}, {Carretero}, {Chang}, {Costanzi}, {da Costa}, {De Vicente}, {Desai}, {Diehl}, {Doel}, {Eifler}, {Everett}, {Evrard}, {Flaugher}, {Frieman}, {Garc{\'\i}a-Bellido}, {Gerdes}, {Gruendl}, {Gschwend}, {Gutierrez}, {Hartley}, {Hollowood}, {Huterer}, {James}, {Krause}, {Kuehn}, {Kuropatkin}, {Maia}, {March}, {Marshall}, {Menanteau}, {Miquel}, {Ogando}, {Paz-Chinch{\'o}n}, {Romer}, {Roodman}, {Sanchez}, {Scarpine}, {Serrano}, {Sevilla-Noarbe}, {Smith}, {Suchyta}, {Tarle}, {Thomas}, {Tucker}, {Walker}, \& {DES Collaboration}}]{Tanoglidis_21}
{Tanoglidis}, D., {Drlica-Wagner}, A., {Wei}, K., {et~al.} 2021, \apjs, 252, 18

\bibitem[{{Toloba} {et~al.}(2023){Toloba}, {Sales}, {Lim}, {Peng}, {Guhathakurta}, {Roediger}, {Wang}, {Mihos}, {C{\^o}t{\'e}}, {Durrell}, \& {Ferrarese}}]{Toloba_23}
{Toloba}, E., {Sales}, L.~V., {Lim}, S., {et~al.} 2023, \apj, 951, 77

\bibitem[{{Trujillo} {et~al.}(2019){Trujillo}, {Beasley}, {Borlaff}, {Carrasco}, {Di Cintio}, {Filho}, {Monelli}, {Montes}, {Rom{\'a}n}, {Ruiz-Lara}, {S{\'a}nchez Almeida}, {Valls-Gabaud}, \& {Vazdekis}}]{Trujillo_19}
{Trujillo}, I., {Beasley}, M.~A., {Borlaff}, A., {et~al.} 2019, \mnras, 486, 1192

\bibitem[{{Tsatsi} {et~al.}(2017){Tsatsi}, {Lyubenova}, {van de Ven}, {Chang}, {Aguerri}, {Falc{\'o}n-Barroso}, \& {Macci{\`o}}}]{Tsatsi_17}
{Tsatsi}, A., {Lyubenova}, M., {van de Ven}, G., {et~al.} 2017, \aap, 606, A62

\bibitem[{{Valenzuela} \& {Remus}(2024)}]{Valenzuela_Remus_24}
{Valenzuela}, L.~M. \& {Remus}, R.-S. 2024, \aap, 686, A182

\bibitem[{{Valenzuela} {et~al.}(2024){Valenzuela}, {Remus}, {Dolag}, \& {Seidel}}]{Valenzuela_24}
{Valenzuela}, L.~M., {Remus}, R.-S., {Dolag}, K., \& {Seidel}, B.~A. 2024, \aap, 690, A206

\bibitem[{{van Dokkum} {et~al.}(2019){van Dokkum}, {Danieli}, {Abraham}, {Conroy}, \& {Romanowsky}}]{vanDokkum_19b}
{van Dokkum}, P., {Danieli}, S., {Abraham}, R., {Conroy}, C., \& {Romanowsky}, A.~J. 2019, \apjl, 874, L5

\bibitem[{{van Dokkum} {et~al.}(2018{\natexlab{a}}){van Dokkum}, {Danieli}, {Cohen}, {Merritt}, {Romanowsky}, {Abraham}, {Brodie}, {Conroy}, {Lokhorst}, {Mowla}, {O'Sullivan}, \& {Zhang}}]{vanDokkum_18}
{van Dokkum}, P., {Danieli}, S., {Cohen}, Y., {et~al.} 2018{\natexlab{a}}, \nat, 555, 629

\bibitem[{{van Dokkum} {et~al.}(2018{\natexlab{b}}){van Dokkum}, {Danieli}, {Cohen}, {Romanowsky}, \& {Conroy}}]{vanDokkum_18b}
{van Dokkum}, P., {Danieli}, S., {Cohen}, Y., {Romanowsky}, A.~J., \& {Conroy}, C. 2018{\natexlab{b}}, \apjl, 864, L18

\bibitem[{{van Dokkum} {et~al.}(2022{\natexlab{a}}){van Dokkum}, {Shen}, {Keim}, {Trujillo-Gomez}, {Danieli}, {Dutta Chowdhury}, {Abraham}, {Conroy}, {Kruijssen}, {Nagai}, \& {Romanowsky}}]{vanDokkum_22}
{van Dokkum}, P., {Shen}, Z., {Keim}, M.~A., {et~al.} 2022{\natexlab{a}}, \nat, 605, 435

\bibitem[{{van Dokkum} {et~al.}(2022{\natexlab{b}}){van Dokkum}, {Shen}, {Romanowsky}, {Abraham}, {Conroy}, {Danieli}, {Chowdhury}, {Keim}, {Kruijssen}, {Leja}, \& {Trujillo-Gomez}}]{vanDokkum_22b}
{van Dokkum}, P., {Shen}, Z., {Romanowsky}, A.~J., {et~al.} 2022{\natexlab{b}}, \apjl, 940, L9

\bibitem[{{van Dokkum} {et~al.}(2015){van Dokkum}, {Abraham}, {Merritt}, {Zhang}, {Geha}, \& {Conroy}}]{vanDokkum_15}
{van Dokkum}, P.~G., {Abraham}, R., {Merritt}, A., {et~al.} 2015, \apjl, 798, L45

\bibitem[{{Vazdekis} {et~al.}(2015){Vazdekis}, {Coelho}, {Cassisi}, {Ricciardelli}, {Falc{\'o}n-Barroso}, {S{\'a}nchez-Bl{\'a}zquez}, {La Barbera}, {Beasley}, \& {Pietrinferni}}]{Vazdekis_15}
{Vazdekis}, A., {Coelho}, P., {Cassisi}, S., {et~al.} 2015, \mnras, 449, 1177

\bibitem[{{Villaume} {et~al.}(2022){Villaume}, {Romanowsky}, {Brodie}, {van Dokkum}, {Conroy}, {Forbes}, {Danieli}, {Martin}, \& {Matuszewski}}]{Villaume_22}
{Villaume}, A., {Romanowsky}, A.~J., {Brodie}, J., {et~al.} 2022, \apj, 924, 32

\bibitem[{{Wang} \& {Peng}(2023)}]{Wang_23}
{Wang}, B. \& {Peng}, Y. 2023, \apjl, 950, L22

\bibitem[{{Wasserman} {et~al.}(2018){Wasserman}, {Romanowsky}, {Brodie}, {van Dokkum}, {Conroy}, {Abraham}, {Cohen}, \& {Danieli}}]{Wasserman_18}
{Wasserman}, A., {Romanowsky}, A.~J., {Brodie}, J., {et~al.} 2018, \apjl, 863, L15

\bibitem[{{Wolf} {et~al.}(2010){Wolf}, {Martinez}, {Bullock}, {Kaplinghat}, {Geha}, {Mu{\~n}oz}, {Simon}, \& {Avedo}}]{Wolf_10}
{Wolf}, J., {Martinez}, G.~D., {Bullock}, J.~S., {et~al.} 2010, \mnras, 406, 1220

\end{thebibliography}
\begin{appendix}
\section{Spectral resolution characterisation}
\label{appendix:spectral_res}
\begin{figure}
    \centering
    \includegraphics[width=\linewidth]{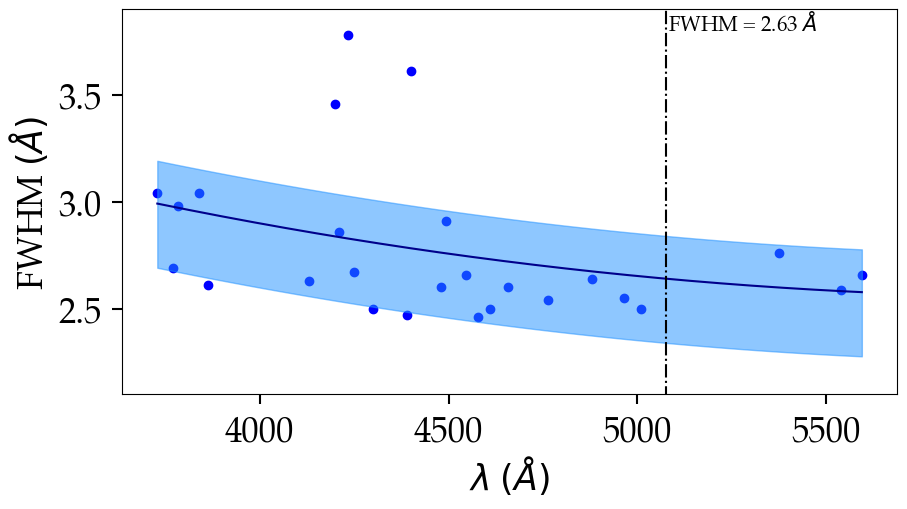}
    \includegraphics[width=\linewidth]{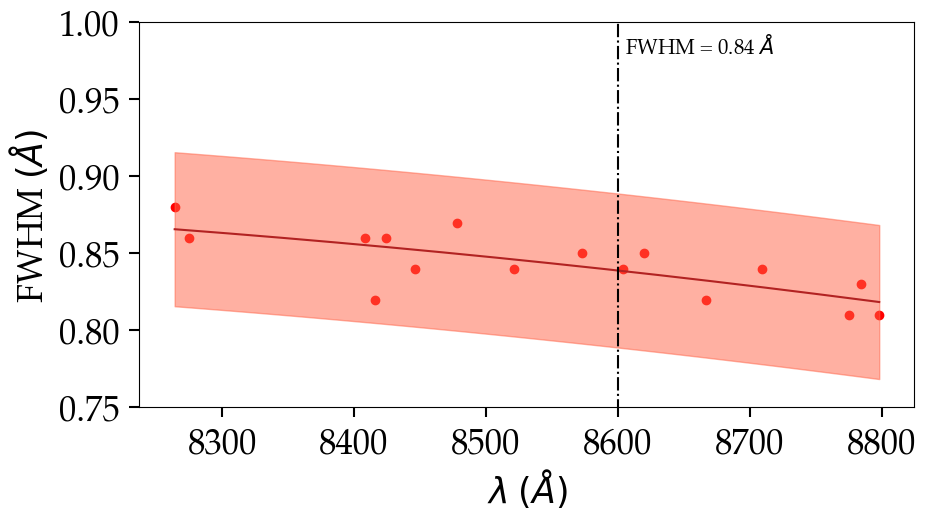}
    \caption{Spectral resolution profiles (line spread function), given by the FWHM as a function of wavelength. The resolution adopted for the blue side is 2.63~\AA\ (R = 1908) at 5075~\AA. For the red side, the resolution is nearly constant across the spectrum with an adopted resolution of 0.84~\AA\ (R = 10238) at 8600~\AA.}
    \label{fig:spectral_resolution}
\end{figure}

The spectral resolution was measured from arc lamp files and is R = 1908 at 5075\,\AA\ ($\sigma_{\text{inst}} = 66.8$\,km\,s$^{-1}$) for the blue arm and R = 10238 at 8600\,\AA\ ($\sigma_{\text{inst}} = 12.4$\,km\,s$^{-1}$) for the red arm. 
The resolution ranges from R = 1390 at 3700\,\AA\ to R = 2113 at 5600\,\AA\ on the blue arm, and from R = 9762 at 8200\,\AA\ to R = 10476 at 8800\,\AA\ on the red arm, as can be seen in Fig~\ref{fig:spectral_resolution}. The median line width is FWHM = 2.65 and 0.84\,\AA\ in the blue and red arms, respectively. Using these minimum and maximum values of resolution on the red side yielded an uncertainty of 1.2 km s$^{-1}$ in the recovered velocity dispersion. This value has been quadratically added to the final uncertainty quoted for the velocity dispersion of FCC~224.

\section{Sky subtraction}
\label{appendix:sky_sub}

\begin{figure*}
    \centering
    \includegraphics[width=0.9\linewidth]{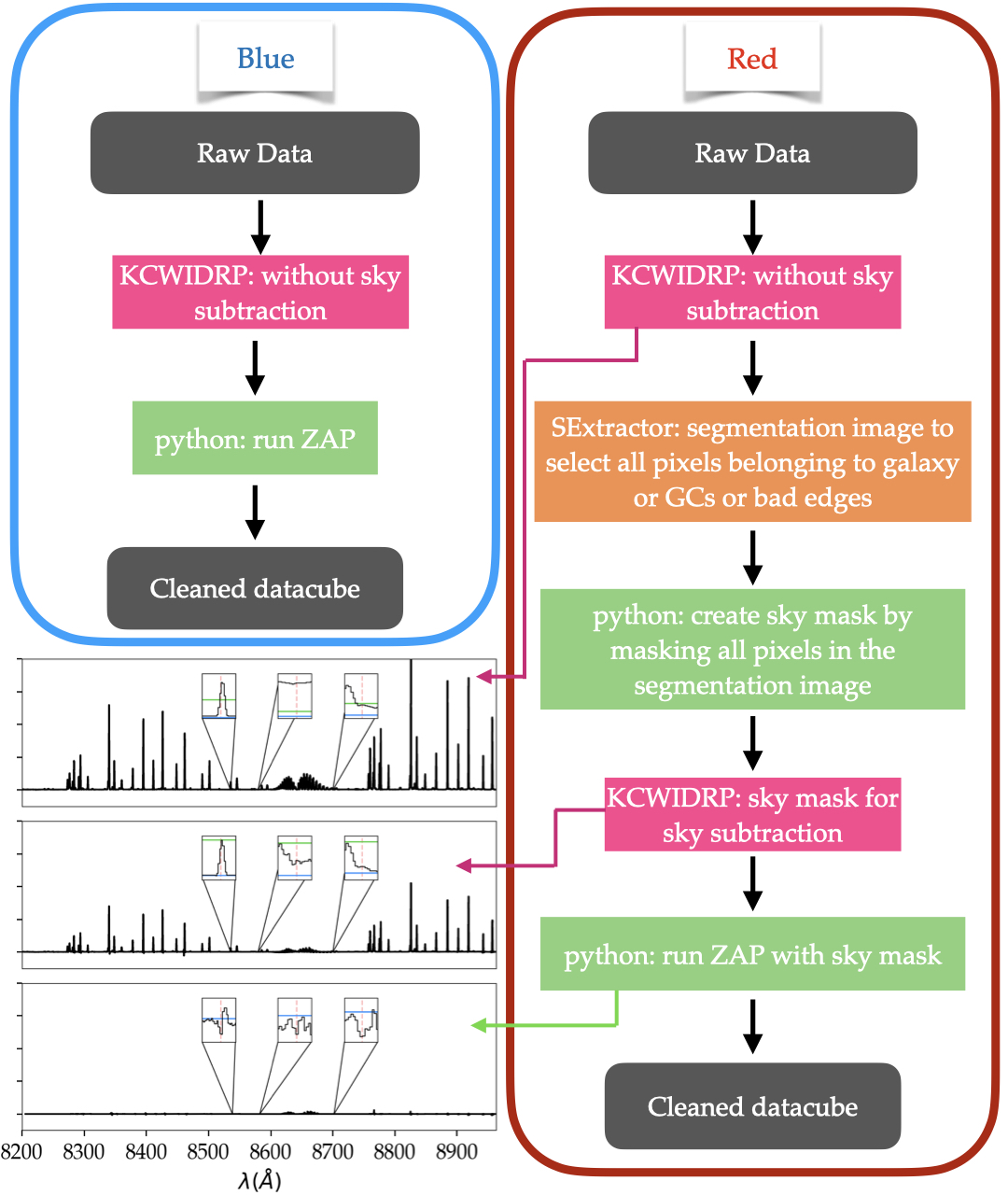}
    \caption{Sky Subtraction methodology. On the top left we show the steps applied to reduce the blue-arm data. The right column shows the steps involved in reducing the red-arm data, much more complex given the cosmic ray frequency and presence of many skylines. The bottom left panel shows the power of each method in removing the contribution of skylines in the red-arm spectra. The blue and green lines in the bottom left panel compare the sky levels at each step of the sky subtraction process. Green is after running only \texttt{KCWIDRP} with a sky mask and the blue shows the sky level after applying \texttt{ZAP}.}
    \label{fig:sky_subtraction}
\end{figure*}

The blue-arm data were reduced with on-chip sky subtraction. For the red arm, additional sky subtraction steps included creating a sky mask with \texttt{SExtractor} \citep{Bertin_02}, and separating the spaxels associated with the galaxy and sky. This sky mask was then used in \texttt{KCWIDRP} for sky subtraction. The algorithm Zurich Atmospheric Purge \citep[ZAP;][]{Soto_16} was applied on residuals in the reduced datacube, using the same sky mask to characterise the sky accurately. The data were collapsed across all spectra associated with both the galaxy and sky, generating integrated spectra. We then used the \texttt{skytweak} function in \texttt{IRAF} with a scaling factor of 1.01 (i.e.\ a 1\% increase) and a shift of 1.3 pixels to correct slight misalignments in the sky and galaxy spectra. This adjustment eliminated residuals in the spectra, allowing clean measurements of key spectral features, including the Ca triplet lines. Fig. \ref{fig:sky_subtraction} summarises all of the reduction steps applied to the blue and red arms of KCWI.

\section{Globular cluster association} 
\label{appendix:gc_association}
\begin{figure}
    \centering
    \includegraphics[width=\linewidth]{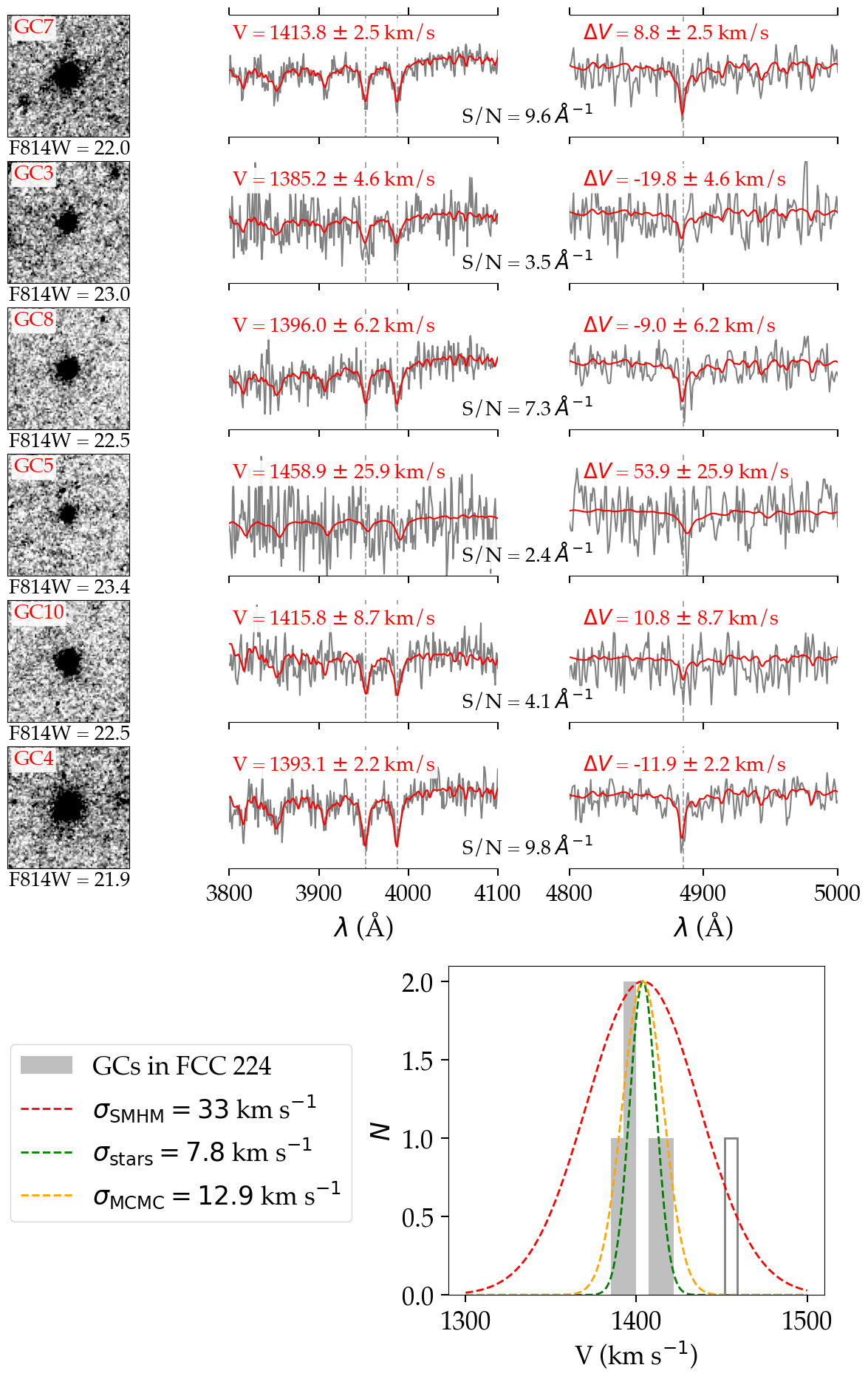}
    \caption{\textit{First six rows:} A close-up view of each GC in the KCWI FoV is shown on the left with their respective F814W magnitudes. On the middle and right columns, the blue-arm KCWI spectrum of each GC is shown, focusing on the CaH+K and H$\beta$ regions. At the top of each spectrum, the recovered radial velocity of the GC is shown, as well as its offset from the galaxy's systemic velocity (determined from the integrated starlight in the blue arm). At the bottom of each spectrum, the S/N is shown. \textit{Last row}: Radial velocity distribution of the GCs around FCC~224. The empty bin is GC5 which was excluded from the analysis because of its low S/N. The three gaussian curves show the expected velocity dispersion if the galaxy follow the stellar mass-halo mass relation \citep{Moster_13} in red, the velocity dispersion that stars alone contribute in green and the result from MCMC of the velocity dispersion of the GC system obtained using the uniform priors in orange.}
    \label{fig:GCs_confirmed}
\end{figure}

The GC candidates around FCC 224 were identified from HST/WFC3 imaging data \citep{Tang_25}. Six GCs were within the FoV of KCWI. To analyse the spectrum of each GC, we select all of the spaxels related to the GC based on the segmentation image, and use an annulus of 3 pixels around each of them to perform a local sky subtraction. We confirm the association of all six GCs through spectroscopic analysis, as can be seen in Fig. \ref{fig:GCs_confirmed}. We then stack all six spectra to study the overall properties of the GC system. The combined stacked spectrum of the GCs has a S/N of 20.1\,\AA$^{-1}$. We find the recessional velocity of the stacked GC spectrum to be $V_{\text{GCs}} = 1408.8 \pm 2.2$\,km s$^{-1}$, i.e., consistent with that of the galaxy. The GC stellar populations also matched the galaxy, with $t_M = 10.1 \pm 0.3$\,Gyr and $[\text{M/H}] = -1.3 \pm 0.2$ dex, supporting a coeval formation scenario. 

\section{Quiescence in low density environments}
\label{appendix_environment}

\subsection{Environment characterisation}
\begin{figure}
    \centering
    \includegraphics[width=\linewidth]{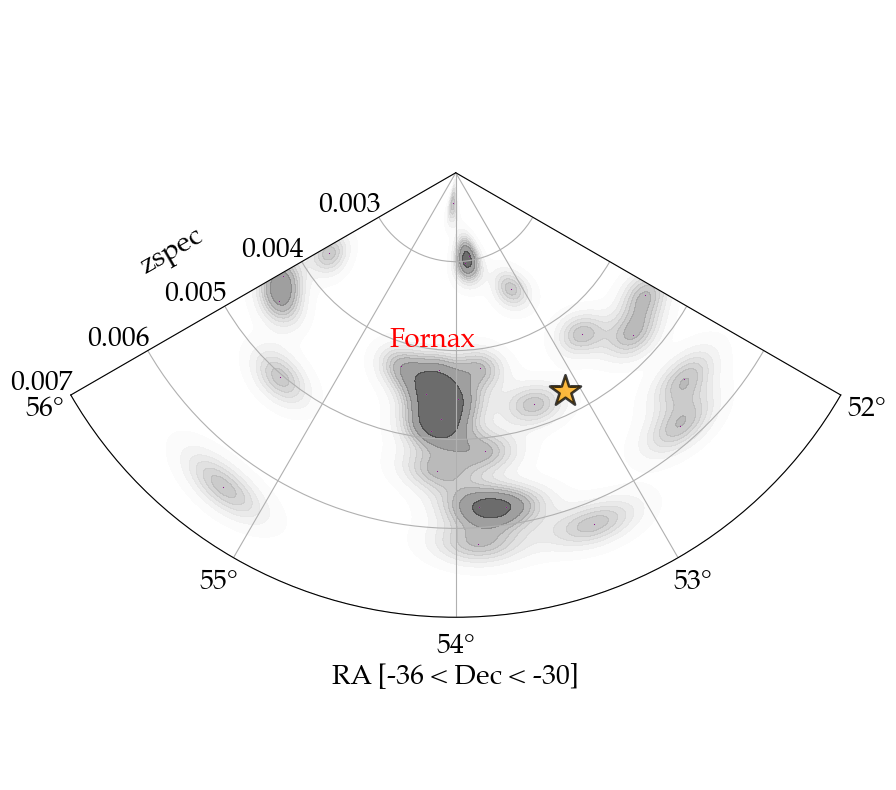}
    \includegraphics[width=\linewidth]{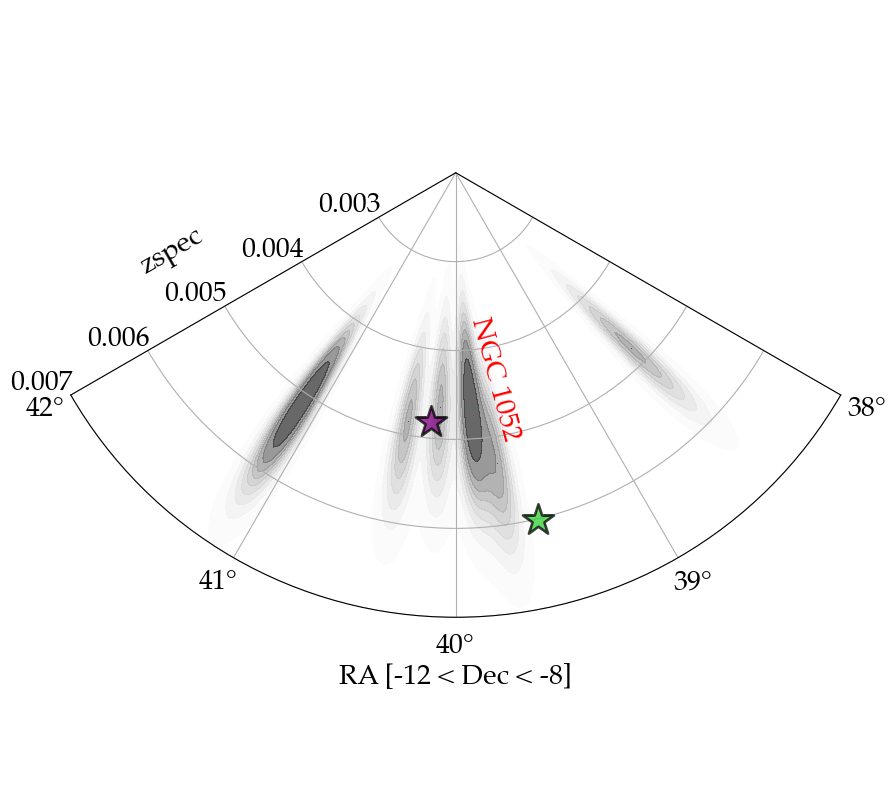}
    \caption{Characterisation of the environments of the galaxies. \textit{Upper panel:} FCC 224 (shown with the orange star marker). \textit{Lower panel:} DF2/DF4 (green and purple stars, respectively). These are based on the 2MASS XSC, revealing that the galaxies are in low-density regions: FCC 224 in the outskirts of the Fornax Cluster and DF2/DF4 in the sparsely populated NGC 1052 group, distant from massive galaxies.}
    \label{fig:environments}
\end{figure}

We used the 2 MASS Redshift Survey \citep[2MRS,][]{Huchra_12} to derive the large scale structure around the galaxies. We have selected only galaxies with magnitudes brighter than 10.5 in the $K_s$ band in the 2MRS survey. The large scale structure maps shown in Fig. \ref{fig:environments} show that FCC~224, as well as DF2 and DF4, are far away from any high-density regions and any massive galaxies, being thus characterised as populating low-density environments.

\subsection{The low fraction of quiescent dwarf galaxies in the outskirts of the Fornax cluster}

To evaluate the rarity of quiescent dwarf galaxies like FCC 224 in low-density environments, we analysed the phase-space distribution and stellar population ages of dwarf galaxies in the Fornax Cluster using data from the SAMI-Fornax dwarf survey \citep{Scott_20, Eftekhari_22, RomeroGomez_23}. This dataset includes detailed measurements of galaxy properties, allowing us to compare the age and spatial distribution of FCC 224 with a well-characterised sample of Fornax dwarfs.

\begin{figure*}
    \centering
    \includegraphics[width=\linewidth]{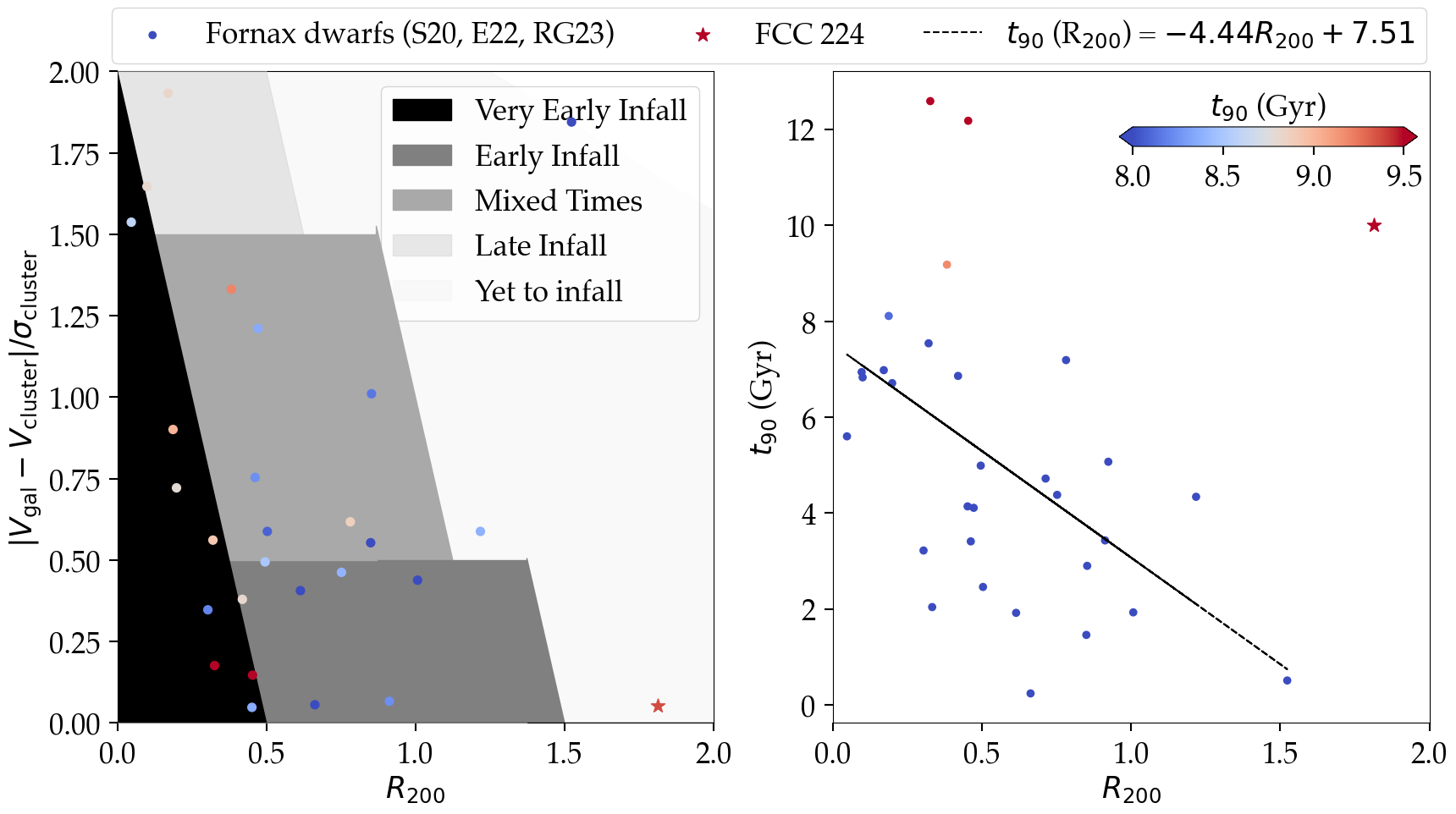}
    \caption{FCC~224 quiescence in the outskirts of the Fornax Cluster. \textit{Left:} Phase-space diagram for FCC~224. The different infall regions from the simulations of Rhee et al. (2017) \protect\cite{Rhee_17} are shown as shaded regions, going from darker to lighter to show galaxies infalling at progressively later times. We find that FCC~224 is in the yet to infall region, being very close in velocity space, but far away spatially. \textit{Right:} Relation between the age ($t_{90}$) and the projected distance to the cluster core. Dwarf galaxies from the SAMI-Fornax dwarf survey are shown for comparison \cite{Scott_20, Eftekhari_22, RomeroGomez_23}. In both panels, the galaxies are colour-coded by their age. The fit shows a clear trend of galaxies becoming younger the furthest away from the cluster center they are located. FCC~224 stands out for being an old galaxy very distant from the cluster core, lying more than 4$\sigma$ from the fit.}
    \label{fig:lowdensity_quiescent}
\end{figure*}

Using the recovered velocity and the coordinates of FCC 224, we placed the galaxy in the phase-space diagram proposed by Rhee et al. (2017) \cite{Rhee_17} as shown in Fig. \ref{fig:lowdensity_quiescent} to see where it lies within the Fornax Cluster. We assume a systemic velocity of the Fornax cluster of V$_{\rm sys}$ = 1425 km s$^{-1}$, $\sigma$ = 374 km s$^{-1}$, and a virial radius of $R_{200} = 2$ deg \citep[= 0.7 Mpc,][]{Raj_20,SmithCastelli_24}. The coordinates of NGC 1399 were used as the centre of the cluster. One can see that although FCC 224 is very close to the centre of the cluster in velocity space, it is far away spatially, lying at nearly $\sim 1.8$ virial radii. According to the simulations of Rhee et al. (2017) \cite{Rhee_17}, FCC 224 is yet to infall the Fornax cluster or has escaped it after a pericentre passage (i.e., a backsplash galaxy; \cite{Benavides_21}).

The right panel of Fig. \ref{fig:lowdensity_quiescent} shows the relationship between galaxy age (represented by $t_{90}$, the time at which 90\% of the stellar mass formed) and distance from the cluster core. We derived this relation using the SAMI-Fornax dwarf sample, which spans a range of environments from the dense core to the cluster outskirts. A clear trend emerges, with galaxies further from the cluster core having younger stellar populations. FCC 224, with its old stellar population ($t_{90} =10.2$ Gyr) and significant spatial distance from the Fornax Cluster core (i.e. 1.8 $R_{200}$), lies more than $4\sigma$ above the fit, making it a clear outlier.

These findings highlight the unusual nature of FCC 224 as a quiescent, old galaxy located in a low-density environment. Such a configuration is exceptionally rare among dwarf galaxies, providing further evidence of FCC 224’s distinct evolutionary pathway.
\end{appendix}
\end{document}